\documentclass[11pt]{article}
\usepackage{amssymb,amsmath,amsthm,graphicx,ulem}
\usepackage{graphicx}
\usepackage{epsfig}
\usepackage{bm} % Include bold math: \bm{} creates bold letters symbols in math mode
\usepackage{amsfonts,amssymb,amsmath}
\usepackage{latexsym}
\usepackage[english]{babel}
\usepackage{subfigure}

\def\beq{\begin{equation}}
\def\eeq{\end{equation}}
\def\bea{\begin{eqnarray}}
\def\eea{\end{eqnarray}}

\newcommand{\eqn}{\begin{eqnarray}}
\newcommand{\eeqn}{\end{eqnarray}}
\newcommand{\arr}{\begin{eqnarray*}}
\newcommand{\earr}{\end{eqnarray*}}

\newcommand{\eg}{{\it e.g.,}\ }
\newcommand{\ie}{{i.e.,}\ } %\newcommand{\ie}{{\it i.e.,}\ }

\newcommand{\lp}{\left(}
\newcommand{\rp}{\right)}

\def\ii{\textrm{i}}

\begin{document}

\setlength{\unitlength}{1mm}

\thispagestyle{empty} \rightline{\small DCPT-10/19}
 \vspace*{2cm}

\begin{center}
{\bf \LARGE  Ultraspinning instability of rotating black holes}\\

%\vspace*{1.5cm}
\vspace*{2.5cm}

 {\bf \'Oscar J.~C.~Dias$^{a}\,$},
{\bf Pau Figueras$^{b}\,$}, \\
{\bf Ricardo Monteiro$^{a,c}\,$}, {\bf Jorge E.~Santos$^{a,c}\,$}
 \vspace*{0.5cm}

{\it $^a\,$ DAMTP, Centre for Mathematical Sciences, University of
Cambridge,\\
Wilberforce Road, Cambridge CB3 0WA,
United Kingdom}\\[.3em]
{\it $^b\,$ Centre for Particle Theory \& Department of Mathematical Sciences, \\
Science Laboratories, South Road, Durham DH1 3LE, UK}\\[.3em]
{\it $^c\,$ Dept. de F\'{\i}sica e Centro de F\'{\i}sica do Porto,
Faculdade de Ci\^encias da Universidade do Porto, Rua do Campo
Alegre 687, 4169 - 007 Porto, Portugal}\\[.3em]

\vspace*{0.5cm} {\tt O.Dias@damtp.cam.ac.uk,
pau.figueras@durham.ac.uk, R.J.F.Monteiro@damtp.cam.ac.uk,
J.E.Santos@damtp.cam.ac.uk}

% \vspace{0.5cm} {June 12, 2009}
\end{center}

\begin{abstract}
Rapidly rotating Myers-Perry black holes in $d\geq 6$ dimensions
were conjectured to be unstable by Emparan and Myers. In a previous
publication, we found numerically the onset of the axisymmetric
ultraspinning instability in the singly-spinning Myers-Perry black
hole in $d=7,8,9$. This threshold signals also a bifurcation to new
branches of axisymmetric solutions with pinched horizons that are
conjectured to connect to the black ring, black Saturn and other
families in the phase diagram of stationary solutions. We firmly
establish that this instability is also present in $d=6$ and in
$d=10,11$. The boundary conditions of the perturbations are
discussed in detail for the first time and we prove that they
preserve the angular velocity and temperature of the original
Myers-Perry black hole. This property is fundamental to establish a
thermodynamic necessary condition for the existence of this
instability in general rotating backgrounds. We also prove a
previous claim that the ultraspinning modes cannot be pure gauge
modes. Finally we find new ultraspinning Gregory-Laflamme
instabilities of rotating black strings and branes that appear
exactly at the critical rotation predicted by the aforementioned
thermodynamic criterium. The latter is a refinement of the
Gubser-Mitra conjecture.
\end{abstract}

\noindent

%\keywords{Kerr/CFT correspondence, Kerr black hole, Entropy}

%\vfill
\vfill \setcounter{page}{0} \setcounter{footnote}{0}
\newpage

\tableofcontents

%\newpage

%%%%%%%%%%%%%%%%%%%%%%%%%%%%%%%%%%%%%%%%%%%%%%%%%%%%%%%%%%%%%%%%%%%%%%%%%%%
%%%%%%%%%%%%%%%%%%%%%%%%%%%%%%%%%%%%%%%%%%%%%%%%%%%%%%%%%%%%%%%%%%%%%%%%%%%

%%%%%%%%%%%%%%%%%%%%%%%%%%%%%%%%%%%%%%%%%%%%%%%%%%%%%%%%%%%%%%%%%%%%%%%%%%%%
\setcounter{equation}{0}
\section{Introduction}
%%%%%%%%%%%%%%%%%%%%%%%%%%%%%%%%%%%%%%%%%%%%%%%%%%%%%%%%%%%%%%%%%%%%%%%%%%%%

The turn of the century witnessed a renewed interest in Einstein's
gravity.  The research focus extends nowadays to numerical
simulations of the full time evolution (inspiral, merger and
ring-down phases) of black hole binary systems
\cite{Pretorius:2005gq}, numerical simulations of high energy
collisions of black holes \cite{Sperhake:2008ga}, and the study of
higher-dimensional black holes \cite{Emparan:2008eg}, mainly in
vacuum and in asymptotically anti-de Sitter (AdS) spacetimes. The
former two endeavours are of the utmost interest for gravitational
wave experiments. On the other hand, the latter programme was
triggered by the quest for a microscopic description of black holes
in a theory of quantum gravity (string theory), by the emergence of
TeV-scale gravity scenarios relevant for the LHC, and by the
realization that black holes describe thermal phases of  gauge
theories in holographic gauge/gravity dualities. The properties of
higher-dimensional black holes, namely their stability, will be the
main focus of the present work, and we review its motivations in
more detail next.

Black hole thermodynamics and the thermal spectrum of Hawking
radiation strongly suggests the existence of a statistical
mechanical description of black holes in terms of some underlying
microscopic degrees of freedom. This microscopic description
necessarily requires quantum gravity. One of the most compelling
candidates for such a theory is string theory, which inevitably
requires gravity in extra dimensions. Furthermore, the observation
that the black hole entropy is proportional to the horizon area
strongly suggests that the quantum degrees of freedom of the black
hole are distributed over a surface \cite{'tHooft:1993gx}. This led
to the formulation of the holographic principle according to which
quantum gravity in a given volume should have a description in terms
of a quantum field theory on its boundary. The last decade witnessed
concrete realizations of this holographic idea, namely the AdS/CFT
duality and more generically the gauge/gravity correspondence
\cite{adsCFTreview}. The application of these ideas to certain
classes of higher-dimensional black holes has already provided a
statistical description of some of their thermodynamic properties.
Successful examples include: i) a statistical counting of the
Bekenstein-Hawking entropy \cite{Strominger:1996sh}, ii) a
microscopic description of Hawking emission in near-extremal black
holes \cite{Callan:1996dv}, iii) a microscopic description of
superradiant emission \cite{Dias:2007nj}, iv) the map between the
Hawking-Page phase transition in asymptotically AdS black holes and
the confinement/deconfinement phase transition in gauge theories
that share some common properties with QCD \cite{Witten:1998qj}, and
v) the identification of the quasinormal modes of oscillation of a
black hole with the thermalization frequencies of the perturbed
holographic field theory \cite{Horowitz:1999jd}, to mention but a
few. Following these early successes, the gauge/gravity
correspondence has matured and is nowadays a field of research on
its own. Mainly due to the fact that this is a weak/strong coupling
duality, it is recognized as a powerful technical tool to understand
not only gravity but also strongly coupled gauge theories. In recent
years this correspondence has been extended to other systems of
interest, namely, the fluid/gravity correspondence
\cite{Policastro:2001yc,Bhattacharyya:2008jc} and more recently the
condensed matter/gravity correspondence \cite{Hartnoll:2009sz}.
These developments indicate that gravitational physics is becoming a
valuable technical tool to understand several other branches of
physics. Given the holographic nature of these correspondences, and
since we are often interested in gauge theories in four dimensions,
it is therefore important to develop our understanding of gravity in
five and higher dimensions.

The LHC will soon be operating at the TeV energy frontier. Although it was designed to find the Higgs boson and study particle physics beyond the Standard Model, there is the possibility that the LHC discovers extra dimensions. Indeed, in recent years various TeV-scale gravity or braneworld scenarios have been proposed according to which extra dimensions might be detected at the energy scales probed by the LHC (see \cite{Cavaglia:2002si} for a review). The motivation for these proposals is the solution of the hierarchy problem, \ie the huge difference between the Planck scale and the
electroweak scale. By proposing the existence of sufficiently ``large" extra dimensions, the fundamental ($d$-dimensional) Planck mass can be of the order of the electroweak energy ($\sim1$ TeV). The most fascinating outcome of these scenarios is the possibility of producing microscopic black holes at the LHC. This could allow for the experimental determination of the fundamental scale of gravity, the number of extra dimensions, and the decay product of higher-dimensional black holes, including the first evidence of Hawking evaporation \cite{Cavaglia:2002si}. This fascinating possiblity provides an extra motivation to undergo the study of higher-dimensional black holes.

Stimulated by the advances in higher-dimensional gravitational
physics, there is an intensive ongoing  program to extend analytical
tools and numerical relativity to higher dimensions. The
Newman-Penrose and Geroch-Held-Penrose formalisms, and the resulting
Petrov classification, which are important to classify solutions and
study their stability, have been extended to $d>4$
\cite{PetrovNPhighdim}. The separability of some class of
perturbation equations and the existence of completely integrable
geodesic equations in some higher dimensional black holes is
possible due to the fact that there are hidden and explicit
symmetries associated with the principal conformal Killing-Yano
tensor \cite{Page:2006ka}. The study of higher-dimensional gravity
is also motivating the development of completely new analytical
tools, like the blackfold approach
\cite{Emparan:2009at,Emparan:2009vd}, to construct perturbative
solutions based on the fact that for higher-dimensional black holes
the horizons can have widely separated length scales.
Simultaneously, the first non-linear numerical codes in higher
dimensions are starting to be developed mainly to study the time
evolution of instabilities in black holes
\cite{Shibata:2009ad,Zilhao:2010sr,Shibata:2010wz}. This is expected
to be an area of active research in the near future. In addition,
the challenges of higher-dimensional black hole physics have
motivated the introduction of new numerical tools, like the spectral
methods  \cite{Monteiro:2009ke,Dias:2009iu,Dias:2010eu}.

Last but not least, another motivation to study higher-dimensional
black holes  is the understanding of these objects {\it per se}.
Indeed, to uncover the mathematical structure of Einstein's gravity
and its solutions one should treat the number of spacetime
dimensions $d$ as a free parameter in the theory. We should be able
to distinguish between the universal properties of the theory and
the dimension dependent ones.

It is known since the 70's that $4d$ black holes are the simplest
gravitational objects in Nature. Hawking's black hole topology
theorem states that a $4d$ black hole must have an event horizon
with spherical topology \cite{hawkingellis}. Together with Hawking's
rigidity theorem \cite{hawkingellis} this led to the proof of the
$4d$ uniqueness theorems \cite{uniqueness}. As a result, a $4d$
vacuum black hole is fully specified by its conserved charges: mass
$M$ and angular momentum $J$. Therefore, there is a unique black
hole in $4d$ Einstein gravity $-$ the Kerr solution $-$ which has an
upper bound on its angular momentum, $J\leq  G M^2$ ($G$ being
Newton's constant). However, are these universal properties of
gravity? Or are some of them dimension dependent?

The last decade has provided a cascade of fascinating answers to
these questions (see \cite{Emparan:2008eg} for a review). The
higher-dimensional arena is populated by new solutions besides the
Myers-Perry (MP)  black hole with $S^{d-2}$ topology
\cite{myersperry} $-$ the (non-trivial) counterpart of the $4d$ Kerr
solution. The simplest black object in $d$ dimensions is a black
string with horizon topology $S^{d-3}\times \mathbb{R}$ or a black
brane with $S^{d-2-n}\times \mathbb{R}^n$, constructed by adding
trivial flat direction(s) to the metric of the Schwarzchild
geometry. The most surprising analytical solution is however the
$5d$ black ring found by Emparan and Reall \cite{Emparan:2001wn}.
This solution is asymptotically flat, has horizon topology
$S^{2}\times S^1$ and in some regions of the parameter space it can
carry the same conserved charges as the singly-spinning Myers-Perry
black hole. Therefore, this solution explicitly demonstrates that
the topology and uniqueness theorems do not generalize
straightforwardly to higher dimensions when rotation is
considered.\footnote{The situation is considerably different for
static solutions. Indeed, it is proven that the
Schwarzschild-Tangherlini black hole is the unique static solution
\cite{Gibbons:2002av}, and moreover this black hole is stable at the
linear mode level \cite{Gibbons:2002pq}.} Indeed, the generalisation
of the black hole topology theorem imposes much weaker restrictions
on the horizon topology \cite{Galloway:2005mf}. In addition, the
$5d$ uniqueness theorems, restricted to geometries with two $U(1)$
Killing isometries, explicitly show that in order to uniquely
specify a black hole solution, one has to fix the conserved charges
\textit{and} other parameters not related to them
\cite{Hollands:2007aj} (see also \cite{Harmark:2004rm} for some
earlier work). Using the complete integrability of the
five-dimensional Einstein's vacuum equations restricted to solutions
with $\mathbb{R}\times U(1)^2$ Killing isometries, it has been also
possible to construct explicitly the black rings with rotation only
along the $S^2$ (which possess conical singularities)
\cite{Mishima:2005id},\footnote{Ref. \cite{Figueras:2005zp} was able
to construct the same solution in ``ring-like'' coordinates  without
using integrability methods.} the regular doubly-spinning black ring
\cite{Pomeransky:2006bd}, and regular asymptotically flat
multi-black hole objects like black Saturns \cite{Elvang:2007rd},
di-rings \cite{Iguchi:2007is} and bi-cycling rings
\cite{Izumi:2007qx}. These solutions provide further examples of the
richness of higher $d$ gravity. In $d>5$, the integrability methods
are not available, but the perturbative blackfold approach of
\cite{Emparan:2009at} has provided evidence for the existence of
black objects with other (non-spherical) topologies
\cite{Emparan:2007wm,Emparan:2009vd}. All these solutions have
multiple $U(1)$ rotational symmetries and need to be organized in a
classification scheme. To provide such a classification we must
address the question of to what extent the rigidity theorem extends
to higher dimensions. In higher dimensions, stationarity only
implies the existence of one rigid rotational symmetry \cite{hiw}.
However all higher-dimensional solutions known exactly have more
than one such symmetry. A fundamental question is whether there are
stationary black holes with less symmetry than those solutions. This
possibility was first raised in \cite{Reall:2002bh}, where the
existence of stationary black holes with a single $U(1)$ was
conjectured. We shall return to this fundamental issue below.

Having these explicit solutions, the next natural question  is whether they are dynamically stable against linear perturbations. Start with the Kerr solution in vacuum. Smarr observed that, as the rotation increases, the horizon Gaussian curvature at the poles starts positive, becomes zero at a critical rotation, and then goes negative before the Kerr bound $J= G M^2$  is reached \cite{Smarr:1973zz}. He further noticed that a similar behaviour occurs in rotating fluid droplets held by surface tension, where this behaviour signals an instability. However, soon after Smarr made this remark, Teukolsky used the Newman-Penrose formalism to find the master equation that governs the gravitational perturbations of the Kerr black hole, which was used by Whiting to show the mode-by-mode stability of the Kerr black hole \cite{Teukolsky:1973ha}.

What about MP black holes? For $d\geq 6$, the  rotation of a
singly-spinning MP can grow unbounded, as already noticed in
\cite{myersperry}. This led Emparan and Myers to propose that, for
sufficiently high rotation, these black holes should become unstable
against what was called the {\it ultraspinning} limit
\cite{Emparan:2003sy}. The natural way to check the presence of this
instability would be to apply a linear perturbation analysis using
the Newman-Penrose formalism, this time for $d\geq 6$.
Unfortunately, although this formalism and the associated
Geroch-Held-Penrose formalism have been extended to higher
dimensions, it is extremely difficult if not impossible to
manipulate the equations to get decoupled master equations for the
perturbations \cite{PetrovNPhighdim}. So, in practice, we cannot use
it yet to treat the perturbation problem analytically.\footnote{Some
subsectors of the perturbations of some classes of MP black holes
have been decoupled \cite{mpperts,Kunduri:2006qa}, but none of them
shows signs of any instability and indeed they do not contain the
kind of perturbations relevant for the ultraspinning instability.}
Given these limitations, Emparan-Myers provided solid heuristic
arguments, that we highlight next, to conjecture the existence of an
instability for a sufficiently large rotation \cite{Emparan:2003sy}.
Take a MP black hole rotating along a single plane for simplicity.
In the \textit{ultraspinning} regime, $a\gg r_+$ (where
$a=\frac{d-2}{2}\frac{J}{M}$, and $r_+$ is the horizon radius), the
black hole horizon flattens out along the plane of rotation and  its
shape can be approximated by that of black disk of radius $a$ and
thickness $r_+$ \cite{Emparan:2003sy}. In fact, one can take a
precise limit in which the MP black hole metric near the rotation
axis reduces to the metric of a static black membrane with horizon
topology $S^{d-4}\times\mathbb{R}^2$, and where the worldvolume
directions of the membrane correspond to the directions along the
original rotation plane \cite{Emparan:2003sy}. This  is a
far-reaching observation since black membranes are afflicted by the
Gregory-Laflamme (GL) instability \cite{Gregory:1993vy}. This led
the authors of \cite{Emparan:2003sy} to conjecture that
ultraspinning MP black holes should be unstable under a
Gregory-Laflamme-type of instability (see subsection
\ref{subsec:ultramotivations} for more details). According to the
arguments of \cite{Emparan:2003sy}, this instability should become
active when $a\sim r_+$, and it provides an effective dynamical
bound on the angular momentum. As often occurs in physical systems,
and in particular in the context of black branes
\cite{Gubser:2001ac}, the stationary threshold of the ultraspinning
instability is expected to signal also a bifurcation point in a
phase diagram of solutions into a new branch of pinched black
objects. New thresholds are expected at higher rotations, from which
additional branches of pinched solutions should bifurcate. These
conjectured stationary solutions preserve the original MP
symmetries, but have spherical horizons distorted by ripples along
the polar direction, and are conjectured to connect to the black
ring and black Saturn families in a phase diagram of stationary
solutions \cite{Emparan:2007wm}.

The main purpose of the present study is to firmly establish the
properties   of the threshold of this ultraspinning instability. A
practical use of the Newman-Penrose formalism in higher dimensions
is not yet available so we will resort to a numerical approach. In a
previous publication, we have already addressed this problem in
$d=7,8,9$ \cite{Dias:2009iu}. The present manuscript, however,
extends the former study in several directions. We present for the
first time the properties of the instability in the six-dimensional
case. This is an important case since $d=6$ is the lowest dimension
where the instability is present, and it was not addressed in
\cite{Dias:2009iu} due to numerical difficulties. Moreover, we
discuss the instability in $d=10$ and $d=11$ dimensions that are
relevant for string theory. We also address the $d=5$ case, in which
the black hole does not have an ultraspinning regime (in the sense
that defined in subsection \ref{subsec:ultrathermo}). Consequently,
no instability that preserves the spatial isometries of the
background can be present, as we confirm numerically. At a more
technical level, we will provide the first detailed discussion of
the boundary conditions of the problem. While doing so we will prove
that the ultraspinning modes preserve the temperature and angular
velocity of the background, an aspect that was not discussed in
\cite{Dias:2009iu}. This is a keypoint to formally establish the
thermodynamic criterium (which is a necessary condition, not a
sufficient one) for the critical rotation above which black holes
can be afflicted by the ultraspinning instability. It can be seen as
a refinement of the Gubser-Mitra conjecture
\cite{Gubser:2000ec,Reall:2001ag}. This thermodynamic criterium was
first proposed in \cite{Dias:2009iu}, and further developed in
\cite{Dias:2010eu}. We will also prove the  claim of
\cite{Dias:2009iu} that the ultraspinning threshold modes that we
find cannot be pure gauge modes.

The ultraspinning instability is not unique to singly-spinning MP
black holes.  Indeed, \cite{Dias:2010eu} found that it is also
present in the case of MP black holes with equal angular momenta
along the $\lfloor (d-1)/2 \rfloor$ rotation planes that are allowed
in $d$ dimensions (here $\lfloor \rfloor$ stands for the smallest
integer part). This occurs in spite of the fact that for these black
holes, contrary to the singly-spinning case, the angular momenta
have an upper bound. The keypoint is that the aforementioned
thermodynamic criterium (subsection 2.2) still allows for an
ultraspinning regime. In the equal spins case and in odd spacetime
dimensions, the geometry is codimension-1, \ie it only depends on
the radial coordinate. This simplifies considerably the analysis of
the perturbed Einstein equations since they reduce to a coupled
system of ODEs. For this reason, the detailed study of the time
dependence of the instability is much simpler, and
\cite{Dias:2010eu} found, as expected, that the modes responsible
for the ultraspinning instability grow exponentially with time. This
fact provides strong confidence to expect that a similar situation
occurs in the singly-spinning case.

The analysis of the ultraspinning instability in the equal spins case \cite{Dias:2010eu} gave the first evidence for the existence of a new family of topologically spherical black hole solutions with only a {\it single} rotational symmetry, \ie solutions that saturate Hawking's rigidity theorem in higher dimensions \cite{hiw}. There is a previous example of a black object with a single rotational symmetry but different horizon topology \cite{Emparan:2009vd}: the ``helical" black rings that can be perturbatively constructed using the blackfold approach \cite{Emparan:2009at}. These developments confirm the conjecture put forward in \cite{Reall:2002bh} and can fairly be seen as one of the most important legacies of the ultrapinning instability studies and of the blackfold approach. More recently \cite{Kunduri:2010vg}, near horizon geometries of extremal black holes have been found with a single rotational symmetry.

The plan of the paper is the following. In section \ref{sec:ultraspin}, we will review the ultraspinning instability conjecture as formulated in \cite{Emparan:2003sy}, together with the refined thermodynamic criterium of \cite{Dias:2009iu,Dias:2010eu}, and we will formulate the problem of finding the modes responsible for the ultraspinning instability as an eigenvalue problem. In section \ref{sec:bcs}, we will discuss the boundary conditions appropriate to the problem at hand and we will prove that the threshold modes searched numerically cannot be pure gauge. In section \ref{sec:results}, we will present our results and finally we will close with a Discussion. An Appendix contains the technical details of the horizon embeddings.

%%%%%%%%%%%%%%%%%%%%%%%%%%%%%%%%%%%%%%%%%%%%%%%%%%%%%%%%%%%%%%%%%%%%%%%%%%%%
\setcounter{equation}{0}
\section{The ultraspinning instability }
\label{sec:ultraspin}
%%%%%%%%%%%%%%%%%%%%%%%%%%%%%%%%%%%%%%%%%%%%%%%%%%%%%%%%%%%%%%%%%%%%%%%%%%%%
In subsection \ref{subsec:ultramotivations}, we will review the work of \cite{Emparan:2003sy} where the conjecture of the ultraspinning instability of singly-spinning MP black holes was first formulated. In subsection \ref{subsec:ultrathermo}, we will review the thermodynamic arguments that led the authors of \cite{Dias:2009iu,Dias:2010eu} to extend the ultraspinning conjecture to other classes of black holes which need not admit an arbitrarily large angular momentum (per unit mass). Finally, in subsection \ref{subsec:ultraeigenvalue}, we will formulate our perturbation problem as an eigenvalue problem.

\subsection{Motivations}
\label{subsec:ultramotivations}

The metric of the $d$-dimensional asymptotically flat  MP black hole spinning in a single plane is given by \cite{myersperry}
\begin{equation}
 \begin{aligned}
ds^2 =& -\frac{\Delta(r)}{\Sigma(r,\theta)}\left( dt+a\sin^2\theta
\,d\phi \right)^2 + \frac{\sin^2\theta}{\Sigma(r,\theta)}\left[
(r^2+a^2)d\phi-a\,dt\right]^2
+\frac{\Sigma(r,\theta)}{\Delta(r)}\,dr^2+
\Sigma(r,\theta)\,d\theta^2 \\
& +r^2\cos^2\theta\, d\Omega^2_{(d-4)}\,,\label{mpbh}
\end{aligned}
\end{equation}
where $d\Omega^2_{(d-4)}$ is the line element of a unit
$(d-4)-$sphere and
\begin{equation}\label{mpbh:aux}
 \Sigma(r,\theta)=r^2+a^2\cos^2\theta\,,\qquad
\Delta(r)=r^2+a^2-\frac{r_m^{d-3}}{r^{d-5}}\,.
\end{equation}
This solution of the Einstein vacuum equations is characterised by
two parameters, namely the mass-radius $r_m$ and the rotation
parameter $a$, \beq r_m^{d-3}=\frac{16\pi GM}{(d-2){\cal
A}_{d-2}}\,,\qquad a=\frac{d-2}{2}\frac{J}{M}\,, \eeq where ${\cal
A}_{d-2} = 2 \,\pi^{(d-1)/2} / \Gamma[(d-1)/2]$ is the volume of a
unit-radius $(d-2)$-sphere,  $G$ denotes Newton's constant and $M$
and $J$ are the ADM mass and angular momentum, respectively. The
(outer) event horizon lies at the largest real root $r=r_+$ of
$\Delta(r)=0$, that is,
\begin{equation}
r_+^2+a^2-\frac{r_m^{d-3}}{r_+^{d-5}}=0\,.
\end{equation}
For $d=4$ a regular horizon exists for all values of $a$ up to the
Kerr bound,  $a=r_m/2$ $(a=GM)$, which corresponds to an extremal
(i.e., zero temperature) black hole with finite size horizon area.
In $d=5$ the situation is somewhat similar: a regular solution
exists for all values of $a$ \text{strictly smaller}  than
$r_m$.\footnote{Notice that for the black ring of
\cite{Emparan:2001wn} the angular momentum is bounded from below,
but otherwise it can be arbitrarily large.} The solution with
$a=r_m$ corresponds to a naked singularity. On the other hand, for
$d\geq 6$, $\Delta(r)$ always has a (single) positive real root for
\textit{all} values of $a$ and therefore there exist black holes
(which are always non-extremal) with arbitrarily large angular
momentum per unit mass. These black holes were dubbed
``ultraspinning'' in \cite{Emparan:2003sy} and they will be the
object of our study in this paper.

In the limit of large angular momentum, these black holes can be
characterised by  two widely separated length scales on the horizon
\cite{Emparan:2003sy}. Let $\ell_\parallel$ denote the
characteristic length scale in the directions parallel to the
rotation plane, and $\ell_\perp$ in the directions perpendicular to
it. For $a\gg r_+$, we have $\ell_\parallel\sim a$ and
$\ell_\perp\sim r_+$, that is, $\ell_\parallel\gg \ell_\perp$.
Therefore, the MP black hole spreads out along the plane of rotation
and it resembles a black membrane.\footnote{Ref.
\cite{Emparan:2003sy} showed that by taking $a\to \infty$, $r_m\to
\infty$ and $\theta \to 0$ while keeping
$\hat{r}_m^{d-5}=r_m^{d-3}/a^2$ and $\sigma=a\,\sin\theta$ fixed,
the singly-spinning MP black hole metric \eqref{mpbh} near the
rotation axis $\theta=0$, reduces to the metric of a black membrane.
In particular, the spatial directions along the worldvolume of the
membrane correspond to the directions along the original plane of
rotation in \eqref{mpbh}. } This observation has far reaching
consequences because black branes are known to be unstable
\cite{Gregory:1993vy}. This led the authors of \cite{Emparan:2003sy}
to conjecture that rapidly rotating MP black holes should be
unstable under Gregory-Laflamme type perturbations that preserve the
symmetries of the $(d-4)$-sphere. More precisely, choosing a
cylindrical basis in polar coordinates $(\sigma= a\sin\theta,\phi)$
on the plane of the membrane, in the $a\to \infty$ limit, the
unstable modes should be of the form \cite{Emparan:2003sy}:
\begin{equation}
 h_{\mu\nu}\sim e^{\Gamma t}\,J_{m}(\kappa\sigma)\,e^{\ii m\phi}\,\tilde{h}_{\mu\nu}(r)\,,
\end{equation}
where $\kappa$ is the wavenumber along the direction parallel to the rotation. In this paper we will only address the axisymmetric case $(m=0)$, for which the radial profile is given by a cylindrical wave $J_{0}(\kappa\sigma)$. By extrapolating these observations to finite (but sufficiently large) $a$, one concludes that there should exist unstable modes that depend only on $r$ and $\theta$ and do not break any of the  symmetries of the background black hole \cite{Emparan:2003sy}.

By considering the thermodynamics of singly-spinning MP black holes, the authors of \cite{Emparan:2003sy} observed that, for $d\geq 6$, these objects exhibit two markedly different behaviors depending on the value of $a$. For instance, consider the temperature of these black holes:
\begin{equation}
 T=\frac{1}{4\pi}\left(\frac{2\,r_+^{d-3}}{r_m^{d-2}}+\frac{d-5}{r_+}\right).
\label{eqn:temperature}
\end{equation}
For fixed mass, $r_+$ is a monotonically decreasing function of $a$.
Therefore,  it follows from \eqref{eqn:temperature} that, starting
from $a=0$, the temperature decreases as $a$ increases, as in the
Kerr black hole case.  However, at
\begin{equation}
\left(\frac{a}{r_+}\right)_{\textrm{mem}}=\sqrt{\frac{d-3}{d-5}}\,\qquad\Rightarrow
\qquad
\left(\frac{a}{r_m}\right)^{d-3}_\textrm{mem}=\frac{d-3}{2(d-4)}\left(\frac{d-3}{d-5}\right)^{(d-5)/2}\,,\label{eqn:mintemp}
\end{equation}
the temperature reaches a minimum and then it starts growing like $\sim r_+^{-1}$ as $a$ increases. This is the kind of behavior of the temperature of a black membrane as the rotation increases. This point was proposed to give an order of magnitude for the appearance of the instability \cite{Emparan:2003sy}. Notice from \eqref{eqn:mintemp} that the membrane-like behavior occurs for $a\gtrsim r_+$ in all dimensions (see the first column in Table \ref{Table:critRot}). The main result of this paper, which builds on \cite{Dias:2009iu}, will be to show that this picture is indeed correct.

\begin{figure}[t]
\centering
\includegraphics[width = 6 cm]{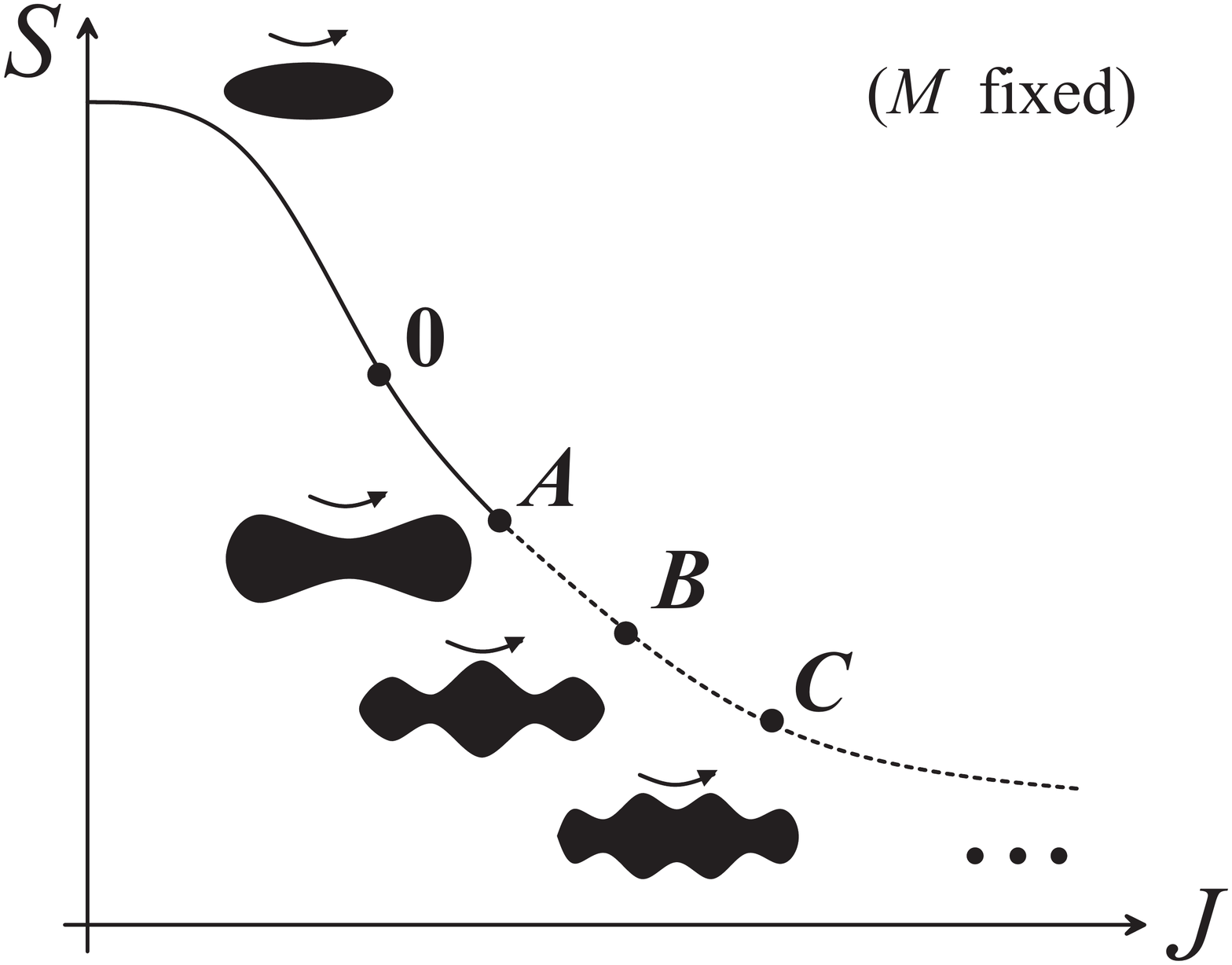}
\caption{\label{fig:phases}Phase diagram of entropy vs.\ angular momentum, at fixed mass, for MP black holes in $d\geq 6$ illustrating the conjecture of \cite{Emparan:2003sy} (see also \cite{Emparan:2007wm}): at sufficiently large spin the MP solution becomes unstable, and at the threshold of the instability a new branch of black holes with a central pinch appear ($A$). As the spin grows new branches of black holes with further axisymmetric pinches ($B,C,\dots$) appear. We determine the points where the new branches appear, but it is not yet known in which directions they run. We also indicate that at the inflection point ($0$), where $\partial^2 S/\partial J^2=0$, there is a stationary perturbation that should not correspond to an instability nor a new branch but rather to a zero-mode that moves the solution along the curve of MP black holes, as discussed in subsection \ref{subsec:ultrathermo}.}
\end{figure}

Furthermore, it is known \cite{Gubser:2001ac} that the threshold zero-mode $(|k|=k_c,\, \Gamma=0)$ of the GL instability gives rise to a new branch of static non-uniform strings. This led the authors of \cite{Emparan:2003sy} to conjecture that the zero-modes of the ultraspinning instability should give rise to a sequence of new branches of stationary and axisymmetric black holes that preserve all the $\mathbb{R}\times U(1)\times SO(d-3)$ isometries of the singly-spinning MP black hole, and that have ripples along the polar direction $\theta$. This conjecture was further refined in \cite{Emparan:2007wm}, which argued that this structure of an infinite sequence of lumpy black holes is in fact needed in order to connect singly-spinning MP black holes to black rings, black Saturns and other black objects which include multiple concentric rings that should exist in $d\geq 6$ (see Fig.\ref{fig:phases}).

%%%%%%%%%%%%%%%%%%%%%%%%%%%%%%%%%%%%%%%%%%%%%%%%%%%%%%%%%%%%%%%%%%%%%%%
\subsection{\label{subsec:ultrathermo}Black hole thermodynamics and the ultraspinning instability conjecture}
%%%%%%%%%%%%%%%%%%%%%%%%%%%%%%%%%%%%%%%%%%%%%%%%%%%%%%

In this subsection, we will refine the arguments of
\cite{Emparan:2003sy} that we have  reviewed above, and we will
present the closely related conjecture made in
Ref.~\cite{Dias:2009iu}. The basic statement of this conjecture is
that classical instabilities associated with stationary zero-modes
can only appear in a regime which we call {\it ultraspinning}. We
analyze this regime for the MP family of solutions. We will also
argue  that a stationary zero-mode of the black hole can correspond:
($i$) to a change in the parameters of the solution, if the
zero-mode can be predicted by the condition of local thermodynamic
stability, or ($ii$) to the threshold of a classical instability of
the black hole, which in our case will correspond to the prediction
of \cite{Emparan:2003sy}.

%%%%%%%%%%%%%%%%%%%%%%%%%%%%%%%%%%%%%%%%%%%%%%%%%%%%%%
\subsubsection{The ultraspinning regime \label{subsubsec:ultraspin}}
%%%%%%%%%%%%%%%%%%%%%%%%%%%%%%%%%%%%%%%%%%%%%%%%%%%%%%

As discussed before, it was conjectured in
Ref.~\cite{Emparan:2003sy} that rapidly-rotating MP black holes with
a single spin in $d\geq6$ were unstable for Gregory-Laflamme-type
modes. An order of magnitude estimate for the critical rotation,
Eq.~(\ref{eqn:mintemp}), was given by considering the thermodynamic
behavior.

The estimate for brane-like behavior is actually a zero-mode of the thermodynamic Hessian
\begin{equation}
\label{thermoHessian} -S_{\alpha\beta}\equiv
-\,\frac{\partial^2{S(x_\gamma)}}{{\partial x_\alpha}{\partial
x_\beta}} \,, \qquad \; x_\alpha=(M,J_i)\,,
\end{equation}
which we readily generalized to accomodate several spins. The condition of local thermodynamic stability is the positivity of this Hessian. It was shown in Ref.~\cite{Dias:2010eu} that $-S_{\alpha\beta}$ possesses at least one negative eigenvalue for any asymptotically flat vacuum black hole (only the Smarr relation and the first law are required). Hence all such black holes are locally thermodynamically unstable. In the MP family, this is the only negative eigenvalue for small rotations. For $d\geq 6$, as the rotation increases, the black hole acquires an additional thermodynamic instability, and a corresponding Gregory-Laflamme type instability of the associated black branes (this is a refinement of the Gubser-Mitra conjecture \cite{Gubser:2000ec}).

Ref.~\cite{Reall:2001ag} (see \cite{Monteiro:2009ke,Dias:2010eu} for
the rotating case)  showed that a zero-mode of the thermodynamic
Hessian is also a zero-mode of the action, i.e. an on-shell
stationary perturbation of the black hole. However, this type of
zero-mode consists of an infinitesimal change in the asymptotic
charges of the black hole solution, within the MP family. In the
singly-spinning case, Eq.~(\ref{eqn:mintemp}) signals a degenerate
point for which an infinitesimal change in the angular momentum at
fixed mass does not change the temperature or the angular velocity
of the MP black hole (indeed, our metric perturbations preserve
these quantities, as we discuss later). This corresponds to the
inflection point $0$ in Fig.~\ref{fig:phases}. Notice that we are
using the MP equation of state $S(M,J_i)$ to identify the zero-mode.
Hence it does not have to be related to the bifurcation to a new
family of solutions, and to the instability commonly associated with
a bifurcation. Only zero-modes of the black hole which are not
associated with the degeneracy of the Hessian (\ref{thermoHessian})
can represent bifurcations and instabilities.

Ref.~\cite{Dias:2009iu} and the present work confirm numerically the conjecture of Ref.~\cite{Emparan:2003sy}, showing that a zero-mode appears at (\ref{eqn:mintemp}), i.e. when the Hessian (\ref{thermoHessian}) is degenerate. More importantly, it is confirmed that additional zero-modes which are not
thermodynamic in origin occur for higher rotations. The latter zero-modes are thresholds for {\it classical} instabilities of the black hole, as explicitly verified for MP codimension-1 solutions in Ref.~\cite{Dias:2010eu}.

Zero-modes which are thermodynamic in origin can be identified as the zero-modes of the simpler reduced Hessian
\begin{equation}
H_{ij}\equiv -\lp{\frac{\partial^2{S}}{{\partial J_i}{\partial
J_j}}}\rp_M= -S_{ij}\,,
\end{equation}
as shown in Ref.~\cite{Dias:2010eu}. In the MP family, the eigenvalues of $H_{ij}$ are all positive for small enough angular momenta, at fixed $M$. However, as some or all of the angular momenta are increased, some eigenvalues of $H_{ij}$ may become negative. We define the boundary of the region in which $H_{ij}$ is positive definite to be the {\it ultraspinning surface}. Following Ref.~\cite{Dias:2009iu}, we shall say that a given black hole is ultraspinning if it lies outside the ultraspinning surface. Hence, as one crosses the ultraspinning surface, the black hole will develop a new thermodynamic instability, and the associated black branes will develop a new classical instability. This is in addition to the usual Gregory-Laflamme instability already present at small angular momenta.

Ref.~\cite{Dias:2009iu} conjectured that classical instabilities whose threshold is a stationary and axisymmetric zero-mode occur only for rotations higher than a thermodynamic zero-mode, i.e. in the ultraspinning regime. We emphasize that the conjecture gives a necessary condition for this instability, not a sufficient one. Notice that by axisymmetric we mean that the $\partial_\phi$ Killing isometry (or $\Omega_i \partial_{\phi_i}$ for more spins) is preserved.

The intuition leading to the conjecture is that modes of lower symmetry are usually the most unstable ones. For instance, the original Gregory-Laflamme instability occurs for the s-wave of the transversal black hole. An additional classical instability will arise after a critical value of the rotation, and it will correspond to a p-wave of the transversal black hole. As the rotation is increased, higher order waves may become unstable. Now, if we consider a black hole, instead of a black brane, the s-wave and the p-wave are associated with the asymptotic charges, mass and angular momenta. Therefore they are associated to purely thermodynamic instabilities. Higher order waves, on the other hand, may become classically unstable as the rotation is increased, starting with the d-wave. Notice that these waves do not affect the asymptotic charges.\footnote{In the codimension-1 case (odd-dimensional equal spins) studied in Ref.~\cite{Dias:2010eu}, where there is a precise harmonic decomposition of perturbations in terms of scalar harmonics of $CP^N$, it was explicitly shown that the first zero-mode changed only the angular momenta, breaking their equality, and that the next zero-modes (associated with classical instabilities/bifurcations) did not affect the asymptotic charges.}

The fact that the thresholds of classical instabilities should be associated with bifurcations to different black hole families highlights the connection between stability and uniqueness. See e.g. Ref.~\cite{Emparan:2007wm}.

%%%%%%%%%%%%%%%%%%%%%%%%%%%%%%%%%%%%%%%%%%%%%%%%%%%%%%%%%%%%%
\subsubsection{Zero-modes of the Myers-Perry family \label{subsubsec:ultraspinMP}}
%%%%%%%%%%%%%%%%%%%%%%%%%%%%%%%%%%%%%%%%%%%%%%%%%%%%%%%%%%%%%

Let us now examine the particular form of $H_{ij}$ in the case of
general MP solutions, with $N= \lfloor (d-1)/2 \rfloor$ spins. We
use the expressions in Ref.~\cite{myersperry}. The horizon area, related to the entropy by $S=A/4$, is
\beq A = \frac{{\cal A}_{d-2}}{r_+^{1-\epsilon}}\, \prod_{i=1}^N
(r_+^2 + a_i^2) \,, \eeq where $\epsilon=0,1$ for odd and even $d$
respectively. The surface gravity $\kappa$, related to the
temperature by $T=\kappa/2\pi$, and the angular velocities on the
horizon $\Omega_i$ are given by \beq \kappa = r_+\, \sum_{i=1}^N
\frac{1}{r_+^2 + a_i^2} -\frac{2-\epsilon}{2\,r_+}\,,\,\qquad
\Omega_i = \frac{a_i}{r_+^2 + a_i^2} \,. \eeq The asymptotic
charges, the mass $M$ and the angular momenta $J_i$, which uniquely
specify a solution, are given by \beq M = \frac{(d-2){\cal
A}_{d-2}}{16 \pi}\,r_m^{d-3} \,,\,\qquad J_i = \frac{2}{d-2}\, a_i\,
M \,. \eeq The reduced thermodynamic Hessian can be explicitly
derived,
\begin{equation}
\begin{aligned}
H_{ij} = \frac{(d-2)\pi}{M \kappa} \left\{
\frac{r_+^2-a_i^2}{(r_+^2+a_i^2)^2} \; \delta_{ij} + 2 \;
\frac{\Omega_i \Omega_j}{\kappa} \left[ \frac{r_+}{(r_+^2+a_i^2)^2}
+ \frac{r_+}{(r_+^2+a_j^2)^2} - \frac{1}{2r_+}+
\frac{\tilde{\Omega}^2}{\kappa} \right] \right\}\,,
\end{aligned}
\end{equation}
where there is no sum over $i,j$, and we have defined $\tilde{\Omega}^2 \equiv \sum_i \Omega_i^2$. The matrix is positive definite in the static case, $a_i=0\, \forall i$, and also in (non-extremal) $d=4,5$.

In the singly-spinning case, say $a\equiv a_1 \neq 0$, we have
\begin{equation}
\begin{aligned}
&H_{11} = \frac{2(d-2)(d-3)\pi\,r_+(r_+^2+a^2)}{M\,[(d-3)r_+^2+(d-5)a^2]^3} \,[(d-3)r_+^2-(d-5)a^2]\,,  \\
&H_{ij} = \frac{(d-2)\pi}{M \kappa\, r_+^2} \delta_{ij} \qquad
\textrm{for} \quad (i,j)\neq(1,1) \,.
\end{aligned}
\label{eqn:hessian1spin}
\end{equation}
There is a single zero-mode in $d\geq6$ which occurs precisely for (\ref{eqn:mintemp}). The associated eigenvector is $\delta_{i1}$, so that the angular momenta which vanish in the background solution are not excited by the perturbation, i.e. the zero-mode keeps the black
hole within the singly-spinning MP family. The ultraspinning regime occurs for rotations higher than Eq.~(\ref{eqn:mintemp}). Ref.~\cite{Dias:2009iu} and the present work find classical instabilities whose thresholds occur in that regime only.

In the equal spins case, whose analysis is also simple, we have
\begin{eqnarray}
H_{ij} = \frac{(d-2)\pi}{M \kappa} \left\{
\frac{r_+^2-a^2}{(r_+^2+a^2)^2} \; \delta_{ij} + 2 \;
\frac{\Omega^2}{\kappa} \left[ \frac{2 r_+}{(r_+^2+a^2)^2} -
\frac{1}{2r_+}+ \frac{n \Omega^2}{\kappa} \right]\,Q_{ij}
\right\}\,,
\end{eqnarray}
where $Q_{ij} =1\; \forall i,j$. An eigenvector $V_i$ of $H_{ij}$ must then be an eigenvector of $Q_{ij}$, which leaves only two options: the eigenvector is such that $V_i=V\, \forall i$, or is such that $\sum_i V_i=0$. In the former case, there can be no zero-mode, since this would require
\begin{eqnarray}
\frac{r_+^2-a^2}{(r_+^2+a^2)^2} + 2 \; \frac{\Omega^2}{\kappa}
\left[ \frac{2 r_+}{(r_+^2+a^2)^2} - \frac{1}{2r_+}+ \frac{N
\Omega^2}{\kappa} \right] N=0\,,
\end{eqnarray}
which can be simplified to $a^2 = - (d-3)r_+^2/(2-\epsilon)$, and thus has no solution for $a$. Hence no instability occurs for modes that preserve the equality between the spins, which is consistent with the results of \cite{Dias:2010eu,Kleihaus:2007dg}. However, the $N-1$ eigenvalues associated with $\sum_i V_i=0$ do change sign once at $|a|=r_+$.\footnote{Notice that Ref.~\cite{Dias:2010eu} used a different radial variable, related to the variable presently used by $\tilde{r}^2 = r^2+a^2$, so that the ultraspinning surface is at $|a|=\tilde{r}_+/\sqrt{2}$.} Therefore, an ultraspinning region exists even though the angular momenta are bounded by extremality, and indeed a classical instability was found in $d=9$ \cite{Dias:2010eu}.

%%%%%%%%%%%%%%%%%%%%%%%%%%%%%%%%%%%%%%%%%%%%%%%%%%%%%%%%%%%%%
\subsection{The eigenvalue problem}
\label{subsec:ultraeigenvalue}
%%%%%%%%%%%%%%%%%%%%%%%%%%%%%%%%%%%%%%%%%%%%%%%%%%%%%%%%%%%%%

According to the preceding discussion, we are interested in modes
that preserve the  $\mathbb{R}\times U(1)\times SO(d-3)$ symmetries
of the background MP black hole \eqref{mpbh}, depending only on the
radial and polar coordinates, $r$ and $\theta$. Thus we consider the
following general ansatz for the perturbed metric
$h_{\mu\nu}$:\footnote{This ansatz  is equivalent to the one
presented in \cite{Dias:2009iu}, but we found that this new form was
better suited for the numerics (see subsection \ref{sec:ttcond}).}
\begin{equation}
\begin{aligned}
ds^2 =& -\frac{\Delta(r)}{\Sigma(r,\theta)}\,e^{\delta\nu_0}\left[
dt+a\sin^2\theta \,e^{\delta\omega}\,d\phi \right]^2 +
\frac{\sin^2\theta}{\Sigma(r,\theta)}\,e^{\delta\nu_1}\left[
(r^2+a^2)d\phi-a\, e^{-\delta\omega}\,dt\right]^2 \\
&
+\frac{\Sigma(r,\theta)}{\Delta(r)}\,e^{\delta\mu_0}\left[dr+\delta\chi\,
\sin\theta \,d\theta \right]^2+
\Sigma(r,\theta)\,e^{\delta\mu_1}\,d\theta^2
+r^2\cos^2\theta\,e^{\delta\Phi}\,
d\Omega^2_{(d-4)}\,,\label{ansatz}
\end{aligned}
\end{equation}
where
$\{\delta\nu_0,\delta\nu_1,\delta\mu_0,\delta\mu_1,\delta\omega,\delta\chi,\delta\Phi\}$
are  small quantities that describe our perturbations and they are
functions of  $(r,\theta)$ only.  Unfortunately a decoupled master
equation that governs the gravitational perturbations, analogous to
the Teukolsky equation \cite{Teukolsky:1973ha} in the case of the
Kerr black hole, is not known for \eqref{mpbh}. Therefore,  in this
paper we will solve numerically the coupled partial differential
equations (PDEs) that govern the class of perturbations we are
interested in. Choosing the traceless-transverse (TT) gauge,
\begin{equation}
h^\mu_{\phantom{\mu}\mu}=0\, \qquad \hbox{and} \quad  \nabla^\mu
h_{\mu\nu}=0\,, \label{TT}
\end{equation}
the variation of the Ricci tensor in vacuum gives the following equations of motion:
\begin{equation} (\triangle_L h)_{\mu\nu}\equiv - \nabla_\rho \nabla^\rho h_{\mu\nu} -2\,
R_{\mu\phantom{\rho}\nu}^{\phantom{\mu}\rho\phantom{\nu}\sigma}
h_{\rho\sigma}=0\,,
\label{Lichnerowicz}
\end{equation}
where $\triangle_L$ is the Lichnerowicz operator. Following \cite{Dias:2009iu} (see also \cite{Dias:2010eu}), we will consider a more general eigenvalue problem
\begin{equation} \label{eigenh}
(\triangle_L h)_{\mu\nu} =-k_c^2 h_{\mu\nu}\,.
\end{equation}
This problem arises when one considers the stability of a uniform rotating black string,
\begin{equation}
ds_{\textrm{string}}=g_{\mu\nu}\,dx^\mu dx^\nu+dz^2\,,
\end{equation}
under perturbations of the form,
\begin{equation}
ds_{\textrm{string}}\to ds_{\textrm{string}}+e^{\ii k_c z
}h_{\mu\nu}(x)dx^\mu dx^\nu\,,
\end{equation}
where $g_{\mu\nu}$ is the metric of the singly-spinning MP black hole \eqref{mpbh}. The same problem arises when one considers the quadratic quantum corrections to the gravitational partition function in the saddle point approximation \cite{Gross:1982cv} (see \cite{Monteiro:2009ke} for the application to the Kerr-AdS black hole.)

Our reason to consider \eqref{eigenh} instead of trying to solve directly \eqref{Lichnerowicz} is that \textit{Mathematica} has very powerful built-in routines to solve generalized eigenvalue problems of this form. Thus, our approach will be to look for solutions of \eqref{eigenh} (which will generically have $k_c\neq 0$), and vary the rotation parameter  $a$ until we find a zero-mode, i.e., a mode with $k_c =0$. Therefore, the solutions with $k_c \neq 0$ will correspond to new kinds of GL instabilities and inhomogeneous phases of ultraspinning black strings (see also \cite{Kleihaus:2007dg}). On the other hand, the $k_c=0$ modes will correspond to asymptotically flat vacuum black holes with deformed horizons of the kind conjectured in \cite{Emparan:2003sy,Emparan:2007wm}, as shown in Fig.~\ref{fig:phases}.

Finally, notice that the ansatz in \eqref{ansatz} is the most general one that respects  the isometries of the background MP black hole and is preserved under diffeormorphisms that depend only on $(r,\theta)$. Ultimately, this is necessary and sufficient to guarantee that \eqref{eigenh} forms a closed system of equations.\footnote{\label{foot:timedep}For the same reasoning,  if we were considering time dependent non-axisymmetric perturbations (that preserve the transverse $(d-4)$-sphere) we would also need to excite metric components of the type $h_{tr}$, $h_{t\theta}$, $h_{\phi r}$, $h_{\phi\theta}$, {\it etc}. We leave this very interesting problem for future work.}

%%%%%%%%%%%%%%%%%%%%%%%%%%%%%%%%%%%%%%%%%%%%%%%%%%%%%%%%%%%%%%%%%%%%%%%%%%%%
\setcounter{equation}{0}
\section{Boundary conditions and gauge fixing}
\label{sec:bcs}
%%%%%%%%%%%%%%%%%%%%%%%%%%%%%%%%%%%%%%%%%%%%%%%%%%%%%%%%%%%%%%%%%%%%%%%%%%%%

In the following subsections we will discuss in detail the boundary
conditions  that we need to impose on the metric perturbations in
order to solve \eqref{eigenh}. In the present situation, we have to
specify boundary conditions at the horizon, $r=r_+$, at asymptotic
infinity, $r\to\infty$, at the rotation axis $\theta=0$, and at the
equator $\theta=\pi/2$. In the following, we shall discuss the
appropriate boundary conditions at the boundaries of the integration
domain. In the final part of this section we show that the threshold
stationary axismmetric zero-modes obeying these boundary conditions
cannot be pure gauge.

%%%%%%%%%%%%%%%%%%%%%%%%%%%%%%%%%%%%%%%%%%%%%%%%%%%%%%%%%%%%%%%%%%%%%%%%%%%%
\subsection{Boundary conditions at the horizon}
%%%%%%%%%%%%%%%%%%%%%%%%%%%%%%%%%%%%%%%%%%%%%%%%%%%%%%%%%%%%%%%%%%%%%%%%%%%%

We shall demand regularity of the metric perturbations on the (future event) horizon ${\cal H}^+$ by imposing regularity of  the Euclideanised perturbed geometry  on ${\cal H}^+$.\footnote{This is equivalent to transforming the metric into regular coordinates  on ${\cal H}^+$ (e.g., ingoing Eddington-Finkelstein coordinates) and requiring that metric perturbations are finite on $\mathcal{H}^+$ in the new coordinate system.} This approach allows us to discuss more straightforwardly the perturbations in the temperature and in the angular velocity.

For $r\approx r_+$, we can write $\Delta(r)=\Delta'(r_+)(r-r_+)+O[(r-r_+)^2]$ with $\Delta'(r_+)>0$,\footnote{Recall that for $d\geq 6$ the singly-spinning MP black hole cannot be extremal and therefore $\Delta'(r_+)>0$ holds for all values of the rotation parameter $a$. The $d=5$ case is special because there is a bound on the rotation whose saturation gives a nakedly singular solution. Therefore we shall only consider non-extremal solutions with $\Delta'(r_+)>0$.} so that  the near horizon geometry of \eqref{mpbh} reads
\begin{eqnarray}
\label{mpbhHorizonAUX}
ds^2&\approx& -\frac{\Sigma\left(r_+,\theta \right) \Delta'\left(r_+ \right) \left(r-r_+\right)}{\left(r_+^2+a^2 \right)^2}\,dt^2 +\frac{\Sigma \left(r_+,\theta \right)}{\Delta'\left( r_+ \right)\left(r-r_+\right)}\,dr^2 \\
&& + \Sigma\left(r_+,\theta \right)d\theta^2+\frac{\left(r_+^2+a^2\right)^2 \sin^2\theta }{\Sigma\left(r_+,\theta \right)}\left(d\phi-\frac{a}{r_+^2+a^2}dt\right)^2+r_+^2\cos^2\theta\, d\Omega^2_{(d-4)}\,.\nonumber
\end{eqnarray}
This suggests the introduction of a new azimuthal coordinate
\begin{equation}
\widetilde{\phi}=\phi-\Omega_H t\,,\qquad
\Omega_H=\frac{a}{r_+^2+a^2}\,, \label{AngVel}
\end{equation}
with $\tilde\phi\sim \tilde\phi+2\pi$. Wick-rotating the time coordinate,
\begin{equation}
t=-\ii \,\widetilde{\tau}\,,\qquad \widetilde{\tau}=\frac{\tau}{2\pi T_H} \qquad \hbox{with} \quad T_H=\frac{\Delta'(r_+)}{4\pi\lp r_+^2+a^2 \rp}\,, \qquad \hbox{and}\quad \widetilde \tau\sim \widetilde\tau+\frac{1}{T_H}\,,
\end{equation}
 and  defining a new radial coordinate $\rho$ as
\begin{equation}  \qquad
 r= r_+ +\frac{\Delta'(r_+)}{4} \rho^2\,,
\end{equation}
 the Euclidean near horizon $(r\approx r_+)$ geometry of the background solution \eqref{mpbh} can be written in a manifestly regular form:
 \begin{equation}
 \label{mpbhHorizon}
ds^2 \approx
 \Sigma\left(r_+,\theta \right)\left[ \rho^2d\tau^2 +d\rho^2 \right]+ \Sigma\left(r_+,\theta \right)d\theta^2+\frac{\left(r_+^2+a^2\right)^2 \sin^2\theta }{\Sigma\left(r_+,\theta \right)}\,d\widetilde{\phi}^2 +r_+^2\cos^2\theta\,d\Omega^2_{(d-4)}\,.
\end{equation}
Indeed, at the $\partial_\phi$ axis of rotation ($\theta=0$) we have
an explicitly regular $S^2$ with no conical singularity. Moreover,
the polar coordinate singularity at $\rho=0$ can be removed by a
standard coordinate transformation into Cartesian coordinates
$(x,y)$. Note that a conical singularity at $\rho=0$ is avoided
because we have chosen the period of the original Euclidean time
coordinate $\tilde{\tau}$ to be $\beta=1/T_H$. To sum up, regularity
at the horizon of the background solution requires that we identify
$(\widetilde{\tau},\phi)\sim(\widetilde{\tau},\phi+2\pi)\sim(\widetilde{\tau}+\beta,\phi-\ii\,\Omega_H\beta)$.
Furthermore, this procedure  identifies $\Omega_H$ with the angular
velocity of the black hole and $T_H$ with its temperature.

The boundary conditions for the metric perturbations can now be determined demanding that $h_{\mu\nu} dx^\mu dx^\nu$ is a regular symmetric 2-tensor when expressed in coordinates where the background metric is regular. To do this, introduce manifestly regular 1-forms,
\beq
\label{SmoothForm1}
E^{\tau}=\rho^2 d\tau=x\,dy-y\,dx\,,\qquad E^{\rho}=\rho\,
d\rho=x\,dx+y\,dy\,.
 \eeq
In terms of these 1-forms, the metric perturbation near the horizon reads
\begin{equation}
\begin{aligned}
h_{\mu\nu}\,dx^\mu\,dx^\nu &\approx \Sigma\left(r_+,\theta\right)
\left[ \delta\nu_0-2 a^2 \sin^2\theta
\left(\frac{\Sigma\left(r_+,\theta\right) \Delta'(r_+) +2 r_+
\left(r_+^2+a^2\right)}{\Sigma^2\left(r_+,\theta
\right)\Delta'\left(r_+\right)}\right)\delta\omega \right]
\rho^2d\tau^2
\\
&~ +\Sigma\left(r_+,\theta \right) \delta\mu_0\, d\rho^2
+\frac{4\,\Sigma\left(r_+,\theta \right)\sin \theta }
{\Delta'(r_+)}\,\frac{\delta\chi}{\rho^2}\, E^{\rho} d\theta
 -\frac{4 \ii \, a\left(r_+^2+a^2\right)^2 \sin^2\theta}{\Sigma\left(r_+,\theta \right)\Delta'\left(r_+\right)}
 \, \frac{ \delta\omega}{\rho^2} E^{\tau} d\widetilde{\phi}
 \\
&~  +\Sigma\left(r_+,\theta \right) \delta\mu_1\, d\theta^2
+\frac{\left(r_+^2+a^2\right)^2 \sin^2\theta}
{\Sigma(r_+,\theta)}\,\delta\nu_1\, d\widetilde{\phi}^2 +r_+^2\cos^2\theta^2\,
\delta\Phi\, d\Omega_{d-4}^2 \,.
\end{aligned}
\label{eqn:Regularhmetric:r+}
\end{equation}
Regularity of the perturbed geometry then requires that
\begin{equation}
\delta\chi,\,\delta\omega =O(\rho^2)\,, \qquad \delta\nu_0-\delta\mu_0=O(\rho^2)\,, \qquad \partial_\rho\delta\mu_1,\,\partial_\rho\delta\nu_1,\,\partial_\rho\delta\Phi=O(\rho)\,,
\label{eqn:BC:r+}
\end{equation}
near $\rho=0$.

It is important to note that we have imposed regularity of both, and
{\it separately},   the background metric $\bar{g}_{\mu\nu}$ and the
perturbed metric $h_{\mu\nu}$. This implies that perturbations
obeying \eqref{eqn:BC:r+} preserve the angular velocity and
temperature of the background black hole.\footnote{Notice that we
could have chosen different boundary conditions by requiring
regularity of the full metric
$g_{\mu\nu}=\bar{g}_{\mu\nu}+h_{\mu\nu}$. Perturbations obeying such
boundary conditions would generically change the temperature and
angular velocity of the background geometry and we shall not
consider them here, but would necessarily lead to a singular
$h_{\mu\nu}$ seen as a two-tensor on the background $g_{\mu\nu}$.}

The regularity analysis of the boundary conditions is not complete without checking that the boundary conditions \eqref{eqn:BC:r+} are consistent both with the eigenvalue Lichnerowicz equations \eqref{Lichnerowicz} and with the TT gauge conditions \eqref{TT}. We have explicitly confirmed this consistency. That is, the first term in the series expansion of the eigenvalue Lichnerowicz equations vanishes after we impose \eqref{eqn:BC:r+}. On the other hand, we can use the TT gauge conditions to express \eg $\{\delta\nu_0,\delta\nu_1,\delta\Phi\}$ as  functions of $\{\delta\mu_0,\delta\mu_1,\delta\omega,\delta\chi\}$ and their first derivatives. Again, the first term of a series expansion of these TT gauge conditions is consistent with \eqref{eqn:BC:r+}.

%%%%%%%%%%%%%%%%%%%%%%%%%%%%%%%%%%%%%%%%%%%%%%%%%%%%%%%%%%%%%%%%%%%%%%%%%%%%
\subsection{Boundary conditions at the axis of rotation}
%%%%%%%%%%%%%%%%%%%%%%%%%%%%%%%%%%%%%%%%%%%%%%%%%%%%%%%%%%%%%%%%%%%%%%%%%%%%

We follow the same strategy as in the previous subsection and we first discuss the regularity of the unperturbed background geometry \eqref{mpbh} at the axis of rotation $(\theta=0)$,  where $\partial_\phi$ vanishes. In the region near $\theta=0$, the background geometry \eqref{mpbh} reads
\begin{equation}
\label{mpbhAxisPhi}
ds^2 \approx -\frac{\Delta(r)}{r^2+a^2}\,dt^2+2 a\lp1-\frac{\Delta(r)}{r^2+a^2}\rp dt d\phi+\left(r^2+a^2\right)\left(d\theta^2+\theta^2 d\phi^2\right)+\frac{r^2+a^2}{\Delta(r)}dr^2+r^2 d\Omega_{(d-4)}^2\,.
\end{equation}
This metric can then be cast in manifestly regular form by changing
to standard Cartesian coordinates $(x,y)$ in the $\theta-\phi$
plane.

To find the boundary conditions that the metric perturbations must satisfy at $\theta=0$ we require that  $h_{\mu\nu} dx^\mu dx^\nu$ is a regular symmetric 2-tensor when expressed in coordinates where the background metric is regular. Introducing the manifestly regular 1-forms,
 \beq
 \label{SmoothForm2}
E^{\theta}=\theta\, d\theta=x\,dx+y\,dy\,,\qquad E^{\phi}=\theta^2 d\phi=x\,dy-y\,dx\,,
 \eeq
the metric perturbation reads
\begin{equation}
\begin{aligned}
h_{\mu\nu}\,dx^\mu\,dx^\nu&\approx
-\frac{\Delta(r)}{r^2+a^2}\,\delta\nu_0\, dt^2+\frac{r^2+a^2}
{\Delta(r)}\,\delta\mu_0\,dr^2 +\frac{r^2+a^2}{\Delta(r)}\,
\delta\chi \,E^{\theta} dr \\
&~~~
+\frac{2a}{r^2+a^2}\left[\left(r^2+a^2+\Delta(r)\right)\delta\omega
-\left(r^2+a^2\right)\delta\nu_1+\Delta(r)\delta\nu_0 \right] E^{\phi}dt\\
&~~~ +\left(r^2+a^2\right)\left[\delta\mu_1 d\theta^2
+\delta\nu_1\theta^2 d\phi^2\right] +r^2\, \delta \Phi
\,d\Omega_{(d-4)}^2 \,.
\end{aligned}
\label{eqn:Regularhmetric:x1}
\end{equation}
Regularity of the perturbed geometry then implies
\begin{equation}
\delta\nu_1-\delta\mu_1=O(\theta^2)\,,\quad\partial_\theta \delta\chi,\,\partial_\theta \delta\omega,\,\partial_\theta\delta\mu_0,\,\partial_\theta\delta\nu_0,\,\partial_\theta \delta\Phi  = O(\theta)\,.
\label{eqn:BC:x1}
\end{equation}
Again we have explicitly checked that the boundary conditions \eqref{eqn:BC:x1} are consistent both with the eigenvalue Lichnerowicz equations \eqref{Lichnerowicz} and with the TT gauge conditions \eqref{TT}.

%%%%%%%%%%%%%%%%%%%%%%%%%%%%%%%%%%%%%%%%%%%%%%%%%%%%%%%%%%%%%%%%%%%%%%%%%%%%
\subsection{Boundary conditions at the equator}
%%%%%%%%%%%%%%%%%%%%%%%%%%%%%%%%%%%%%%%%%%%%%%%%%%%%%%%%%%%%%%%%%%%%%%%%%%%%

Introducing a new polar coordinate $x=\cos\theta$, we find that the metric \eqref{mpbhAxisSphere} near $x= 0$ $(\theta= \pi/2)$ is given by
\begin{equation}
\label{mpbhAxisSphere}
\begin{aligned}
ds^2 &\approx
-\frac{\Delta(r)-a^2}{r^2}\,dt^2-\frac{2a}{r^2}\left[r^2+a^2
-\Delta(r)\right]dt d\phi+\frac{\left(r^2+a^2\right)^2-a^2
\Delta(r)}{r^2}\,d\phi^2\\
&~~~ +\frac{r^2}{\Delta(r)}\,dr^2+r^2\left(dx^2+ x^2
d\Omega_{(d-4)}^2\right)\,.
\end{aligned}
\end{equation}
Once more, this geometry can be put in a manifestly regular form by
changing to Cartesian coordinates.

As in the previous subsections, we now demand that the metric perturbation $h_{\mu\nu} dx^\mu dx^\nu$ is a regular symmetric 2-tensor when expressed in coordinates where the background metric is regular at $x=0$. Introducing the manifestly regular 1-forms,
 \beq
 \label{SmoothForm3}
E^{x}=x\, dx\,,\qquad E^{\Omega}=x^2 d\Omega_{(d-4)}\,,
 \eeq
the metric perturbation near $x=0$ reads
\begin{equation}
\begin{aligned}
h_{\mu\nu}\,dx^\mu\,dx^\nu &\approx
-\frac{\left[\Delta(r)-a^2\right]\delta\nu_0
+2a^2\delta\omega}{r^2}\,dt^2
 +\frac{\left(r^2+a^2\right)^2 \delta\nu_1 -a^2 \Delta(r) \left(\delta\nu_0+2\delta\omega \right)}{r^2}\,d\phi^2  \\
&~~~ +\frac{2a}{r^2}\left[\left(r^2+a^2+\Delta(r)\right)\delta\omega
-\left(r^2+a^2\right)\delta\nu_1+\Delta(r)\delta\nu_0 \right] dtd\phi\\
&~~~
+\frac{r^2}{\Delta(r)}\,\delta\mu_0\,dr^2-\frac{r^2}{\Delta(r)}\left(
\frac{\delta\chi}{x}- \partial_x \delta\chi \right)E^x dr+r^2
\left(\delta\mu_1 \,dx^2+x^2\,\delta\Phi
\,d\Omega_{(d-4)}^2\right)\,.
\end{aligned}
\label{eqn:Regularhmetric:x0}
\end{equation}
Regularity of the perturbed geometry for $x\to 0$ then requires that
\begin{equation}
\begin{aligned}
\delta\chi= O(x)\,, \quad
 \delta\Phi-\delta\mu_1=O(x^2)\,,\quad
 \partial_x\delta\omega,\,
 \partial_x\delta\mu_0,\,
 \partial_x\delta\nu_0,\,
 \partial_x\delta\nu_1=O(x)\,.
\end{aligned}
\label{eqn:BC:x0}
\end{equation}
Again we have explicitly checked that the boundary conditions \eqref{eqn:BC:x0} are consistent both with the eigenvalue Lichnerowicz equations \eqref{Lichnerowicz} and with the TT gauge conditions \eqref{TT}.

%%%%%%%%%%%%%%%%%%%%%%%%%%%%%%%%%%%%%%%%%%%%%%%%%%%%%%%%%%%%%%%%%%%%%%%%%%%%
\subsection{Boundary conditions at the asymptotic infinity}
\label{sec:bcinfinity}
%%%%%%%%%%%%%%%%%%%%%%%%%%%%%%%%%%%%%%%%%%%%%%%%%%%%%%%%%%%%%%%%%%%%%%%%%%%%

At spatial infinity, $r\rightarrow\infty$, we will require that the perturbations preserve the asymptotic flatness of the spacetime. This means that they must decay strictly faster than the background asymptotic solution which, near spatial infinity,  reduces to Minkowski spacetime.\footnote{More precisely, consider a spacetime $(\mathcal M,g)$ which contains a spacelike hypersurface $M_\textrm{ext}$  diffeomorphic to $\mathbb R^n\setminus B(R)$, where $B(R)$ is a coordinate ball of radius $R$. The spacetime is said to be asymptotically flat if the induced metric $h$ on $M_\textrm{ext}$ and the extrinsic curvature $K$ satisfy
\begin{equation}
 h_{ij}-\delta_{ij}=O_k(r^{-\alpha})\,,\qquad K_{ij}=O_{k-1}(r^{-1-\alpha})\,,
\end{equation}
where $r$ is the radial coordinate in $\mathbb R^n$, and we write $f=O_k(r^\beta)$ if $f$ satisfies
\begin{equation}
\partial_{k_1}\ldots \partial_{k_\ell}f=O(r^{\beta-\ell}),\,\qquad 0\leq \ell \leq k\,.
\end{equation}
}

In the asymptotic region, the eigenvalue Lichnerowicz equations \eqref{Lichnerowicz} reduce simply to $\square h_{\mu\nu} \simeq -k_c^2 h_{\mu\nu}$. The regular solutions (there are also irregular solutions that grow exponentially) of these equations decay as
\begin{equation}
\label{eqn:BC:infinity}
h_{\mu\nu}{\bigr|}_{r\rightarrow\infty}\sim \frac {1}{r^\alpha}\,e^{-k_c\,r} \rightarrow 0 \,,
\end{equation}
for some constant $\alpha \geq 0$ that depends on the particular metric component and the number of spacetime dimensions $d$. Therefore, for $k_c\neq 0$ our perturbations decay exponentially to zero at the spatial infinity and hence asymptotic flatteness is guaranteed.

Ultimately one would be interested in studying modes with $k_c=0$,
since these correspond to exact perturbative  solutions to Einstein
\textit{vacuum} equations in $d$ spacetime dimensions. In this case,
one would have to worry about the fall off of the metric
perturbations. However, in our numerical method we will never be
able to find modes for which $k_c=0$ \textit{exactly}, and therefore
the fall off conditions at infinity for these modes are not an issue
for us.

%%%%%%%%%%%%%%%%%%%%%%%%%%%%%%%%%%%%%%%%%%%%%%%%%%%%%%%%%%%%%%%%%%%%%%%%%%%%
\subsection{Imposing the TT gauge conditions and the boundary conditions}
\label{sec:ttcond}
%%%%%%%%%%%%%%%%%%%%%%%%%%%%%%%%%%%%%%%%%%%%%%%%%%%%%%%%%%%%%%%%%%%%%%%%%%%%

We have to solve the Lichnerowicz eigenvalue problem \eqref{eigenh} for  the seven metric perturbations described in \eqref{ansatz}, namely $\{\delta\mu_0,\delta\mu_1,\delta\chi,\delta\omega,\delta\nu_0,\delta\nu_1,\delta\Phi\}$, subject to the TT gauge conditions \eqref{TT}. The gauge conditions allow us to eliminate three functions in terms of the other four and their first derivatives. In this paper we choose to express $\{\delta\nu_0,\delta\nu_1,\delta\Phi\}$ as functions of $\{\delta\mu_0,\delta\mu_1,\delta\chi,\delta\omega\}$ and their first derivatives. Notice that this choice differs from that in \cite{Dias:2009iu}, where the independent variables were chosen to be $\{\delta\mu_0,\delta\mu_1,\delta\chi,\delta\Phi\}$. The reason is that the present choice allowed us to obtain good numerical results in $d=5,6$. As discussed in subsection \ref{sec:bcinfinity}, for $k_c\neq0$  our (regular) perturbations vanish exponentially for $r\to \infty$. However, in addition to the exponential decay, one also has a power law behaviour as in (\ref{eqn:BC:infinity}). This apparently irrelevant extra power law decay seems to make all the difference for the stability and/or accuracy of the numerical code in $d=5,6$. %The key quantity  seems to be $r^2 \delta\Phi$ $-$ see \eqref{ansatz} $-$ which decays as $e^{-|k|\,r}$ in $d=5$ and as $\frac{1}{r}e^{-|k|\,r}$ in $d=6$ (this then introduces a logarithmic behavior once we integrate the perturbation equations). In $d>6$ $r^2 \delta\Phi$ decays faster and does not give rise to any numerical problems. On the other hand, the new independent variable $\delta\omega$ decays exponentially with an additional power of $1/r$ in \textit{all} dimensions. Therefore, the better decay properties at infinty of our new independent functions seems to be key in order to obtain good results in $d=5,6$.
For $d=7,8,9$ we obtain exactly the same results as in \cite{Dias:2009iu}.

To summarize, we solve the gauge conditions \eqref{TT} for $\{\delta\nu_0,\delta\nu_1,\delta\Phi\}$ in terms of $\{\delta\mu_0,\delta\mu_1,\delta\chi,\delta\omega\}$ and their first derivatives. Making these substitutions in the full set of the perturbation equations \eqref{eigenh}, we find that only four equations remain of second order in $\{\delta\mu_0,\delta\mu_1,\delta\chi,\delta\omega\}$. Explicitly, these equations are
\begin{equation}
\begin{aligned}
&(\triangle_L h)_{rr} =-k_c^2 h_{rr}\,,\\
& (\triangle_L
h)_{r\theta} =-k_c^2 h_{r\theta}\,,\\
& (\triangle_L h)_{\theta\theta} =-k_c^2
h_{\theta\theta}\,,\\
& a\,(\triangle_L h)_{tt}+\frac{r^2+a^2+a^2\sin^2\theta}{\lp
r^2+a^2\rp\sin^2\theta}\,(\triangle_L h)_{t\phi}
  + \frac{a}{\lp r^2+a^2\rp\sin^2\theta}\,(\triangle_L h)_{\phi\phi}\\
& \qquad\qquad\qquad\qquad
  =-k_c^2 \left[ a\,h_{tt}+\frac{r^2+a^2+a^2\sin^2\theta}{\lp r^2+a^2\rp\sin^2\theta}\, h_{t\phi}
  + \frac{a}{\lp r^2+a^2\rp\sin^2\theta}\,h_{\phi\phi} \right]\,,
\end{aligned}
\label{eqn:finalsystem}
\end{equation}
and they constitute our final set of equations to be solved.

A non-trivial consistency check of our procedure is to verify that the final equations \eqref{eqn:finalsystem} imply the remaining equations,  which are of third order in the independent variables. We have verified that this is indeed the case.

We will solve numerically the  final eigenvalue problem \eqref{eqn:finalsystem} using spectral methods (see for instance \cite{Trefethen}). In order to do so, we find it convenient to introduce new radial and polar coordinates
\beq
y=1-\frac{r_+}{r}\,,\qquad x=\cos\theta\,,
\eeq
so that $0\leq y \leq 1$ and $0\leq x\leq 1$. The implementation of the method requires less computational power if all functions obey Dirichlet boundary conditions on all boundaries. Therefore, we redefine our independent functions according to
\begin{equation}
\label{eq:qs}
\begin{aligned}
&q_1(y,x) =  \left( 1-\frac{r_+}{r} \right) x(1-x)
\,\delta\mu_0(y,x)\,, \qquad \qquad
q_2(y,x) = r_m^{-1} (1-x)\,\delta\chi(y,x) \,, \\
&q_3(y,x) =  \left( 1-\frac{r_+}{r} \right)x(1-x)
\,\delta\mu_1(y,x)\,,  \qquad \qquad q_4(y,x) =  x
\,\delta\omega(y,x) \,,
\end{aligned}
\end{equation}
so that the $q_i$'s vanish at all  boundaries. This guarantees that the boundary conditions discussed in the previous subsections, equations \eqref{eqn:BC:r+},  \eqref{eqn:BC:infinity}, \eqref{eqn:BC:x0} and \eqref{eqn:BC:x1}, are correctly imposed.

%%%%%%%%%%%%%%%%%%%%%%%%%%%%%%%%%%%%%%%%%%%%%%%%%%%%%%%%%%%%%%%%%%%%%%%%%%%%
%\setcounter{equation}{0}
\subsection{Zero-modes are not pure gauge  \label{sec:NoPureGauge}}
%%%%%%%%%%%%%%%%%%%%%%%%%%%%%%%%%%%%%%%%%%%%%%%%%%%%%%%%%%%%%%%%%%%%%%%%%%%%
In this subsection, we will argue that the TT gauge conditions \eqref{TT} plus our boundary conditions  \eqref{eqn:BC:r+}, \eqref{eqn:BC:infinity}, \eqref{eqn:BC:x0} and \eqref{eqn:BC:x1}, ensure that our modes are physical and not a gauge artifact.

For $k_c>0$, the TT conditions completely fix the gauge since the action of the Lichnerowicz operator on a pure gauge mode is trivial, $\Delta_L \nabla_{(\mu} \xi_{\nu)}=0$. However, TT perturbations with $k_c=0$ can be pure gauge \cite{Gregory:1993vy} and therefore it could be that the stationary perturbation $k_c \to 0$ marking the onset of a new ultraspinning instability is not physical. In the rest of this subsection, we will show that there is no regular pure gauge perturbation that obeys our boundary conditions, \eqref{eqn:BC:r+}, \eqref{eqn:BC:infinity}, \eqref{eqn:BC:x0} and \eqref{eqn:BC:x1}.

Under a gauge transformation with gauge parameter $\xi^\mu$ the metric perturbation transforms as
\begin{equation}
 h_{\mu\nu}\to h_{\mu\nu}+2\,\nabla_{(\mu}\xi_{\nu)}\,. \label{diffeo}
\end{equation}
The most general gauge parameter that preserves our ansatz \eqref{ansatz} is of the form
\begin{equation}
 \xi=\xi_r(r,x)\,dr + \xi_x(r,x)\,dx\,, \label{gaugeparameter}
\end{equation}
where $x=\cos\theta$. We will now prove that such a gauge parameter
$\xi^\mu$  cannot
generate a pure gauge metric perturbation that is regular on all boundaries.

A TT gauge perturbation generated by $\xi^\mu$ must satisfy $\nabla_\mu\xi^\mu=0$ and $\Box\xi^\nu=0$. If we introduce the antisymmetric tensor $F_{\mu\nu}=\nabla_{[\mu}\xi_{\nu]}$ and we consider Ricci flat backgrounds, these conditions reduce to
  \beq
 \partial_\mu\lp \sqrt{-g}\,\xi^\mu \rp=0\,,\qquad \partial_\mu\lp \sqrt{-g}\,F^{\mu\nu}
 \rp=0\,. \label{NoPureGaugeEqs}
 \eeq
Assuming that the solutions of these equations are separable,
  \beq
\xi_r(r,x)=R_r(r)X_r(x)\,,\qquad  \xi_x(r,x)=R_x(r)X_x(x) \,,
\label{separationAnsatz}
 \eeq
the first equation in \eqref{NoPureGaugeEqs} reduces to
  \beq
\partial_r\lp r^{d-4}\Delta R_r \rp = \lambda\, r^{d-4}R_x\,,\qquad
\partial_x\lp x^{d-4}(1-x^2) X_x \rp = -\lambda\, x^{d-4}X_r\,,
\label{separationEq1}
 \eeq
where $\lambda$ is a separation constant. Similarly, a combination of the $r$ and $x$ components of the second equation in
\eqref{NoPureGaugeEqs} yields
  \beq
\partial_r\lp r^{d-4}\Delta \partial_r R_x \rp = \kappa \,r^{d-4}R_x\,,\qquad
\partial_x\lp x^{d-4}(1-x^2) \partial_x X_r \rp = -\kappa\, x^{d-4}X_r\,,
\label{separationEq2}
 \eeq
where $\kappa$ is a second separation constant. Combining the equations \eqref{separationEq1} and \eqref{separationEq2}  we further find that
  \beq
R_r=\frac{\lambda}{\kappa}\,R_x^\prime\,,\qquad
X_x=\frac{\lambda}{\kappa}\,X_r^\prime\,. \label{separationEq3}
 \eeq
We will take equations \eqref{separationEq1} and \eqref{separationEq3} as our independent equations.

Plugging \eqref{separationEq3} into \eqref{separationEq1} gives
  \beq
\partial_r\lp r^{d-4}\Delta \partial_r R_x \rp = \kappa \, r^{d-4}R_x\,,\qquad
 \partial_x\lp x^{d-4}(1-x^2) \partial_x X_r \rp =
-\kappa \, x^{d-4}X_r \,. \label{separationEq4}
 \eeq
 The second of these equations has the solution
  \beq
\small{ X_r(x)= C_1
F\left(\frac{d-3-K}{4},\frac{d-3-K}{4},\frac{d-3}{2} ,x^2\right)+C_2
x^{5-d}
F\left(\frac{7-d-K}{4},\frac{7-d-K}{4},\frac{7-d}{2},x^2\right)\,, }
\label{separationEq5}
 \eeq
where $C_{1,2}$ are integration constants and  we have defined $K\equiv \sqrt{(d-3)^2+4\kappa}$. For $d\geq 6$ (where we find the unstable modes) this solution diverges as $C_2 x^{5-d}$ at $x=0$ unless we set $C_2=0$. In addition,  a $\log(1-x)$ divergence at $x=1$ can only be avoided if $\kappa$ satisfies the following quantization condition:
\begin{equation}
 \kappa=2\ell (d-3+2\ell)\,,\quad\hbox{for}\quad \ell=0,1,2,\ldots\,,\label{eqn:thekappa}
\end{equation}
which follows from the property $\Gamma(-\ell)=\infty$ for non-negative integer $\ell$.

 At this point we just have to solve the first equation in \eqref{separationEq4} with $\kappa$ given by \eqref{eqn:thekappa}. For our purposes it is sufficient to obtain the behavior of the solution near the horizon and near infinity.  Near the horizon, $r\approx r_+$, the most general solution behaves like
  \beq
R_x(r) \approx
 a_0 + a_1 \log\lp 1-\frac{r_+}{r}\rp \qquad \Rightarrow \quad
R_r(r) \approx
 \frac{\lambda a_1}{2\ell(d-3+2\ell)}\frac{1}{r-r_+} \,,
 \label{separationEq6}
 \eeq
where we used \eqref{separationEq3}. However, equation \eqref{diffeo} implies that such a diffeomorphism would generate metric perturbations of the form
  \beq
h_{rr}\approx \frac{\lambda
a_1}{2\ell(d-3+2\ell)}\frac{1}{(r-r_+)^2}\,, \qquad h_{rx}\approx
\frac{\lambda a_1 P(x)}{2(r-r_+)}\,,
\label{separationEq7}
 \eeq
which  diverge at the horizon.  Therefore, our regularity requirements at the horizon force us to set $a_1=0$. Similarly, near infinity a Frobenius analysis gives the asymptotic solution
 \beq
R_x(r)\approx
   b_0 r^{-(d-3+2\ell)}+b_1 r^{2\ell} \,. \label{separationEq8}
 \eeq
The term in this equation proportional to $b_1$ generates, through \eqref{diffeo},  a metric perturbation of the form $h_{rr} \propto b_1 r^{2(\ell-1)}$ near infinity. This perturbation must decay strictly faster than the unperturbed background solution (which at infinity is Minkowski), which implies that we have to impose $b_1=0$ for any $\ell\geq 0$.

Summarizing, the gauge parameter \eqref{gaugeparameter} that could potentially generate a  metric perturbation that is regular both at the horizon and at infinity must have the asymptotic behavior,
  \beq
R_x(r)|_{r\rightarrow r_+} \approx
 a_0 \,, \qquad
R_x(r)|_{r\rightarrow \infty} \approx
 b_0 r^{-(d-3+2\ell)} \,, \label{BC:Rx}
 \eeq

We can now complete our proof. Notice that
\begin{eqnarray}
 0\leq \int_{r_+}^{\infty} dr \,
 r^{d-4}\Delta(r)\left[ \partial_r R_x(r)\right]^2
&=& r^{d-4}\Delta(r) R_x(r) \partial_r
R_x(r){\biggl|}_{r_+}^{\infty} -\int_{r_+}^{\infty}
dr \partial_r \left[ r^{d-4}\Delta(r) \partial_r R_x(r)\right]R_x(r) \, \nonumber\\
 &=& -2\ell(d-3+2\ell)\int_{r_+}^{\infty}r^{d-4}R_x(r)^2 \leq 0 \,, \label{nullRx}
\end{eqnarray}
where we used \eqref{BC:Rx} and \eqref{separationEq2}. But these relations can be satisfied only for $R_x(r)=0$, which in turn implies that $R_r(r)=0$ and hence $\xi^\mu=0$.

Therefore, we have proved that there is no regular gauge parameter
\eqref{gaugeparameter}  that could potentially generate the metric
perturbations that we consider, eq. \eqref{ansatz}. Thus, we
conclude that our regular zero-mode perturbations cannot be pure
gauge.

%%%%%%%%%%%%%%%%%%%%%%%%%%%%%%%%%%%%%%%%%%%%%%%%%%%%%%%%%%%%%%%%%%%%%%%%%%%%
\setcounter{equation}{0}
\section{Results}
\label{sec:results}
%%%%%%%%%%%%%%%%%%%%%%%%%%%%%%%%%%%%%%%%%%%%%%%%%%%%%%%%%%%%%%%%%%%%%%%%%%%%

\begin{figure}
\centering
\includegraphics[width =5.0 cm]{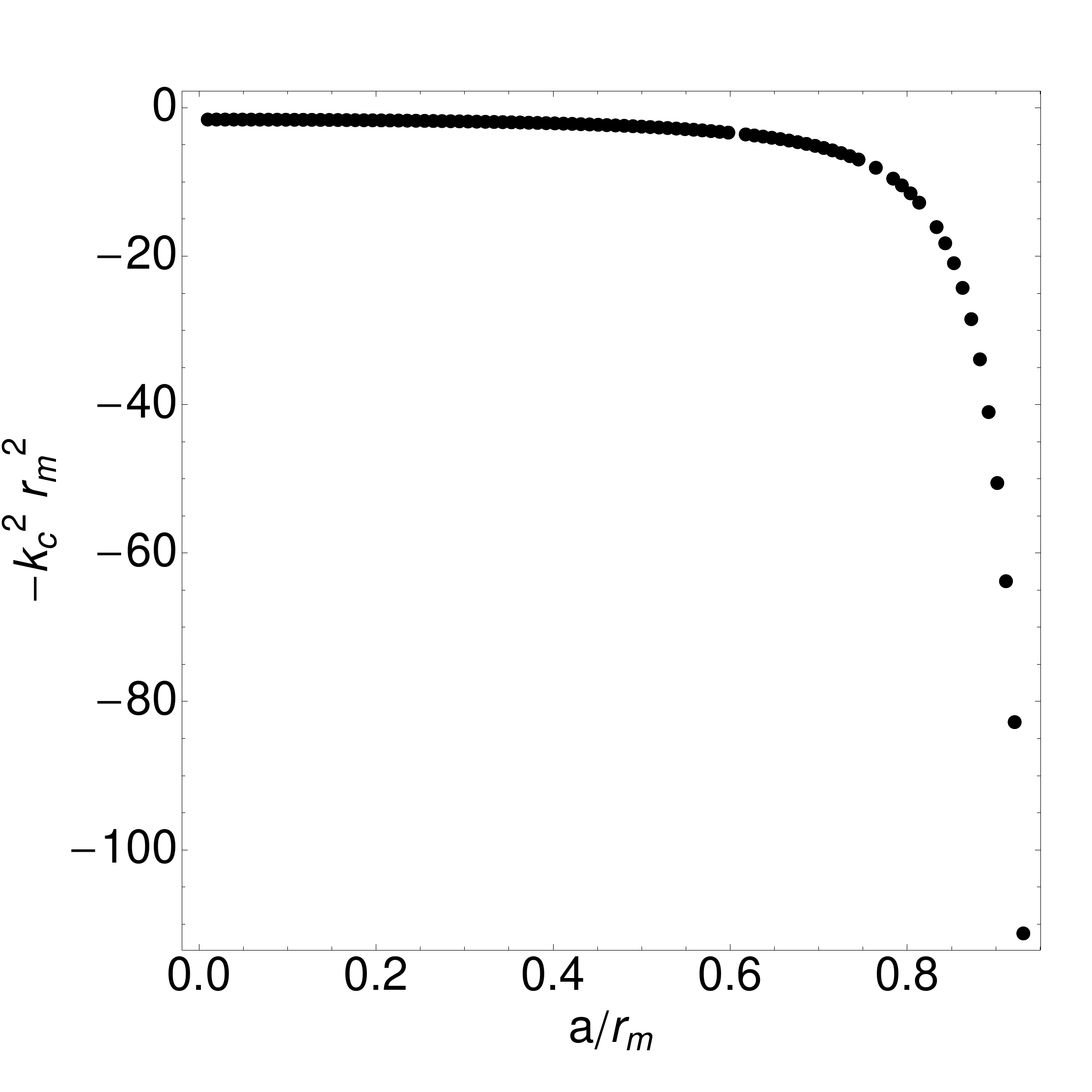}
\hspace{0.5cm}
\includegraphics[width =5.0 cm]{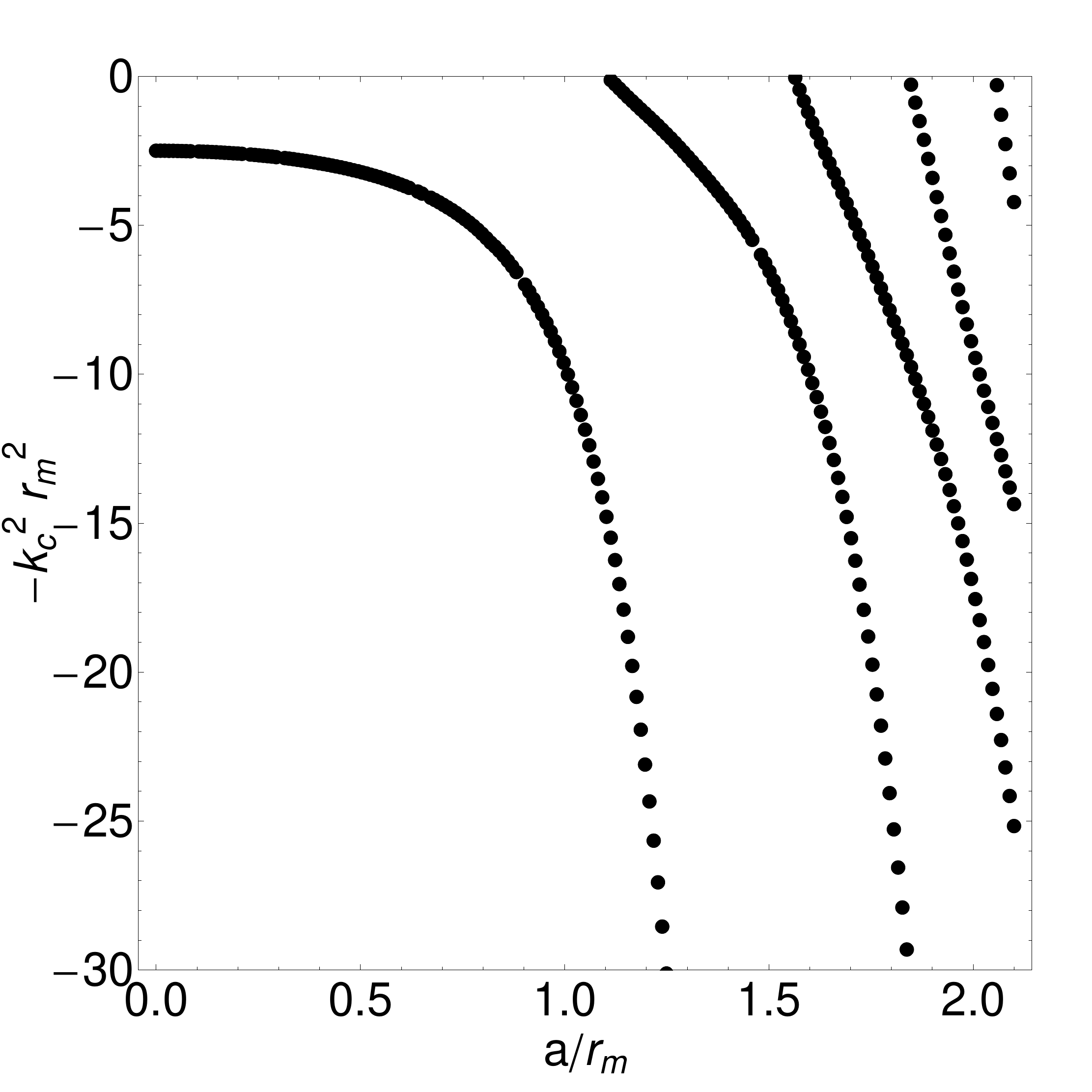}
\hspace{0.5cm}
\includegraphics[width =5.0 cm]{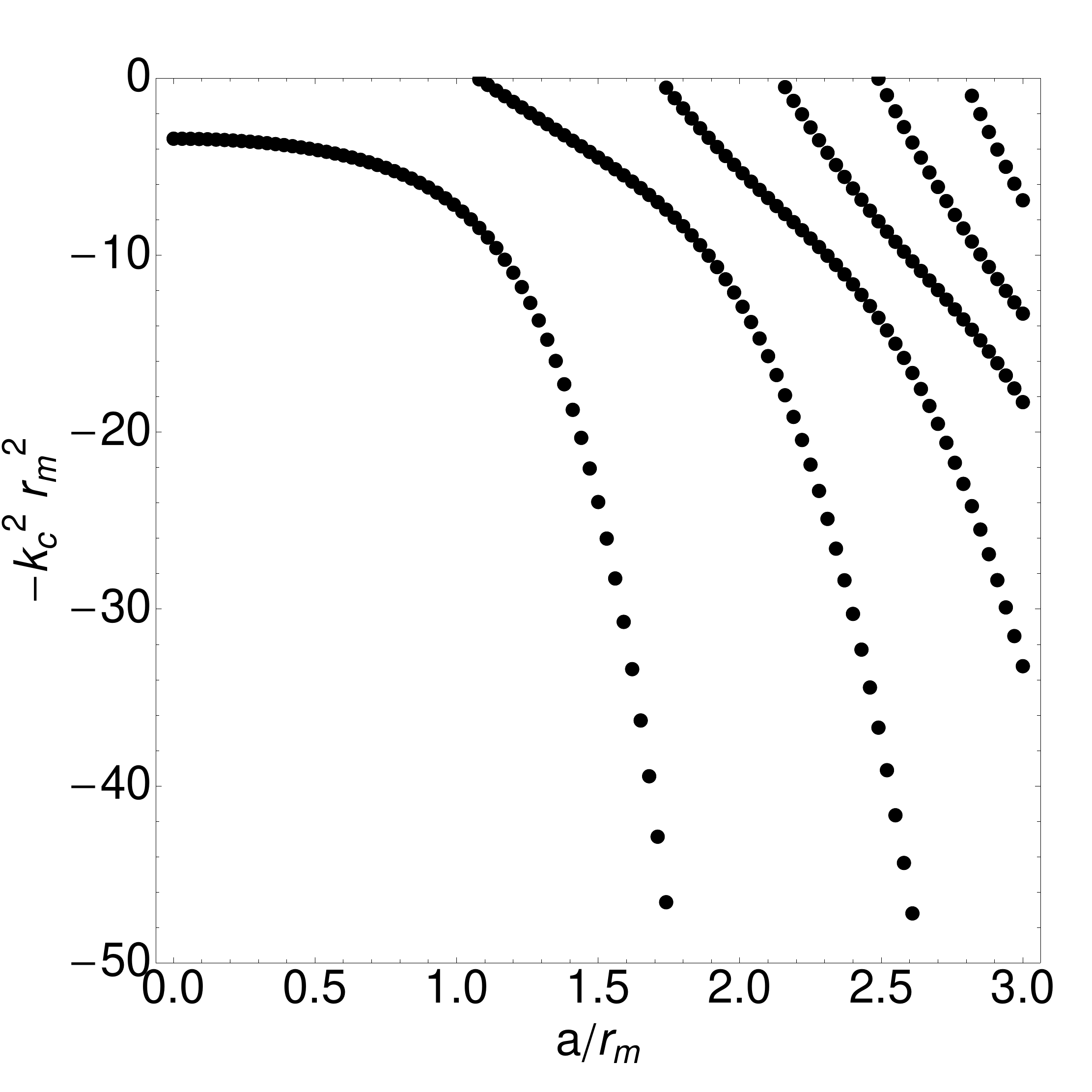}
\caption{\label{fig:negmodes} Negative modes of the singly-spinning
MP black hole in $d=5$ \textit{(left)}, $d=6$ \textit{(centre)}, and
$d=7$ \textit{(right)}.}
\end{figure}

In this section, we present our results for the spectrum of negative
modes of the Lichnerowicz operator.  The actual spectrum in
$d=5,6,7$ is displayed in Fig. \ref{fig:negmodes}; for the other
values of $d$ up to $d=11$ the results are qualitatively similar to
$d=6,7$ and we expect the same to be true for all values of $d$ with
$d>11$. Following \cite{Dias:2009iu},  we plot the dimensionless
negative eigenvalue $-k_c^2r_m^2$ as a function of the
(dimensionless)  rotation parameter $a/r_m$. In Figs.
\ref{fig:perturbations1} and \ref{fig:perturbations2} we show the
actual metric perturbations
$\{\delta\mu_0,\,\delta\mu_1,\,\delta\chi,\,\delta\omega\}$ for
$d=7$, and in Fig.  \ref{fig:embeddings} we display the embeddings
of the unperturbed and the perturbed horizons for the same number of
spacetime dimensions.

%%%%%%%%%%%%%%%%%%%%

\begin{table}[ht]
\begin{eqnarray}
\nonumber
\begin{array}{|c|c|c|c|c|}\hline
 d & (a/r_m)|_{\ell=1} & (a/r_m)|_{\ell=2} &
 (a/r_m)|_{\ell=3} & (a/r_m)|_{\ell=4}\\ \hline\hline
  6  & 1.097  & 1.572 & 1.849 & 2.036\\
  \hline
 7  & 1.075  & 1.714 & 2.141 & 2.487\\
\hline
 8  & 1.061 & 1.770 & 2.275 & 2.725\\
 \hline
9  & 1.051  & 1.792 & 2.337 & 2.807\\
\hline
10 & 1.042 & 1.795 & 2.361 & 2.855 \\
\hline
11 & 1.035 & 1.798 & 2.373 & 2.879 \\
\hline
\end{array}
\end{eqnarray}
\caption{Values of the rotation $a/r_m$ for the first four harmonics of stationary perturbation modes ($k_c=0$). The estimated numerical error is $\pm 3\times 10^{-3}$ in $d=6,\,7$ and $\pm 5\times 10^{-3}$ in $d=8,\,9,\,10,\,11$.} \label{Table:critRot}
\end{table}

%%%%%%%%%%%%%%%%%%%%%%%%%%%%%%%%%%
\subsection{Results for $d=5$}
%%%%%%%%%%%%%%%%%%%%%%%%%%%%%%%%%%

Ref. \cite{Dias:2010eu}  proved that  MP black holes (not
necessarily singly-spinning)  are always locally thermodynamically
unstable. Therefore, it follows from the arguments in subsection
\ref{subsec:ultrathermo} that the spectrum of the Lichnerowicz
operator should admit at least one negative eigenvalue. Since in
$d=5$ the Hessian \eqref{eqn:hessian1spin} is positive definite, the
expectation is that the Lichnerowicz operator should admit one and
only one negative mode. Our results for the spectrum of the
Lichnerowicz operator are depicted in Fig.~\ref{fig:negmodes} (left)
and indeed they confirm this expectation. As the rotation is increased, $k_c$ increases,
in agreement with the results of
\cite{Kleihaus:2007dg,Monteiro:2009ke,Dias:2010eu}. If we interpret
this result from the black string perspective, then more modes
$|k|<k_c$ are GL-unstable, which is expected since the centrifugal
force should make the string ``more unstable''. The negative mode diverges as
$k_c \propto 1/(r_m-a)$, in the singular limit $a\to r_m$.

\subsection{Results for $6\leq d \leq 11$}

\begin{figure}[t!]
\begin{center}
\subfigure[]{
\includegraphics[scale=0.30]{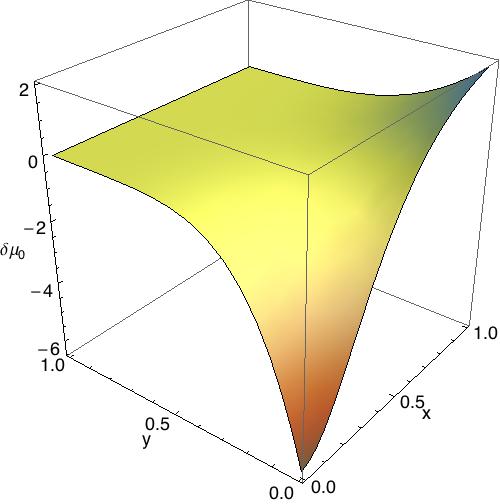}
\label{subfig:mu0l1} } \hspace{0.5cm} \subfigure[]{
\includegraphics[scale=0.30]{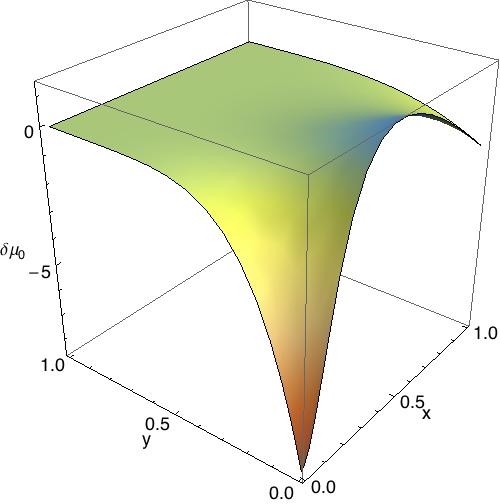}
\label{subfig:mu0l2} } \subfigure[]{
\includegraphics[scale=0.30]{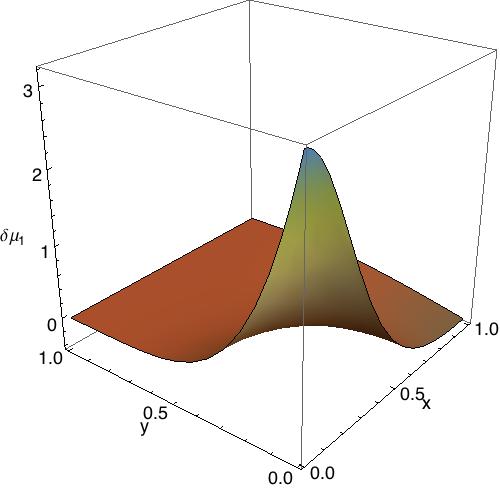}
\label{subfig:mu1l1} } \hspace{0.5cm} \subfigure[]{
\includegraphics[scale=0.30]{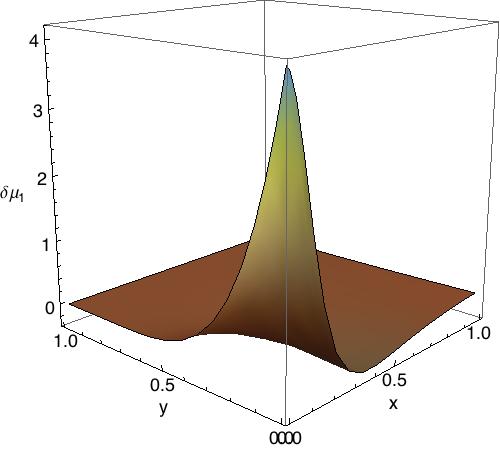}
\label{subfig:mu1l2} }
\end{center}
\caption{Functions $\delta\mu_0(x,y)$ and $\delta\mu_1(x,y)$  for the $\ell=1$ (Figs. \ref{subfig:mu0l1} and \ref{subfig:mu1l1}) and $\ell=2$ (Figs.  \ref{subfig:mu0l2} and \ref{subfig:mu1l2}) modes respectively. The number of zeros at $y=0$ (the horizon) coincides with the integer $\ell$.} \label{fig:perturbations1}
\end{figure}

The situation in $d\geq 6$ is more interesting. The spectrum of the
Lichnerowicz operator in $d=6,7$ is displayed in
Fig.~\ref{fig:negmodes} (center) and (right). We label the different
branches that intersect  $k_c=0$ at finite $a/r_m$ by successive
integers $\ell=1,2,3,\ldots$, and we refer to the corresponding
modes as harmonics, although the equations we are solving do not
seem to separate. The values of $a/r_m$ at which the first few
branches intersect $k_c=0$ are summarised in Table
\ref{Table:critRot}. We note that for each branch the corresponding
integer $\ell$ coincides with the number of zeros that the metric
perturbations $\delta\mu_0(x,y)$ and $\delta\mu_1(x,y)$ have on the
horizon $y=0$ (see Fig. \ref{fig:perturbations1}.)

First, we notice that the Lichnerowicz operator has a negative
eigenvalue  for all values of  $a/r_m$.  This is in agreement with
the thermodynamic argument in subsection \ref{subsec:ultrathermo}.
However, for $d\geq 6$, singly-spinning MP black holes admit an
ultraspinning regime. The ultraspinning surface, which determines
the boundary of this region (in parameter space), is given by the
onset of the membrane-like behavior (see eq. \eqref{eqn:mintemp}):
\begin{equation}
\left(\frac{a}{r_m}\right)^{d-3}_\textrm{mem}=\frac{d-3}{2(d-4)}\left(\frac{d-3}{d-5}\right)^{(d-5)/2}\,.
\label{eqn:acrit}
\end{equation}
According to the thermodynamic argument in subsection
\ref{subsec:ultrathermo}, precisely at  this value of $a/r_m$ the
Lichnerowicz operator should develop a new negative eigenvalue and
we find that this is indeed the case. In the first column of Table
\ref{Table:critRot}, we display our numerical results for the
critical values of $a/r_m$ for the $\ell=1$ mode in various
dimensions. We find that the numerical results agree very well with
the values of $a/r_m$ computed from \eqref{eqn:acrit}. We emphasise
that the $\ell=1$ mode should correspond to a variation of the
parameters within the MP family of solutions such that it preserves
the temperature and angular velocity of the background (this mode is
in correspondence with point $0$ in Fig. \ref{fig:phases}).
Therefore this mode should not correspond to an instability of the
black hole. However, it should give rise to a new type of classical
instability of the associated black string, in which the horizon is
deformed along both the direction of the string and the polar
direction of the sphere.

\begin{figure}[t!]
\begin{center}
\subfigure[]{
\includegraphics[scale=0.30]{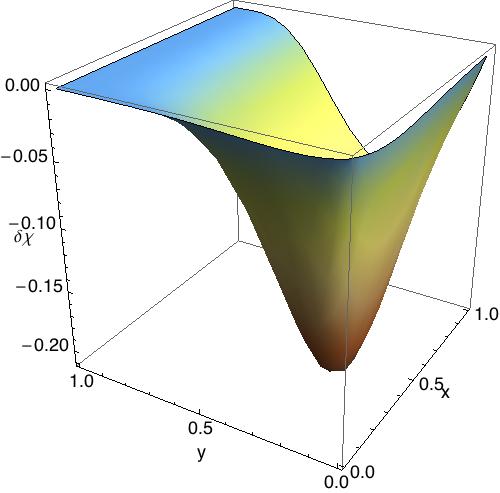}
\label{subfig:chil1} } \hspace{0.5cm} \subfigure[]{
\includegraphics[scale=0.30]{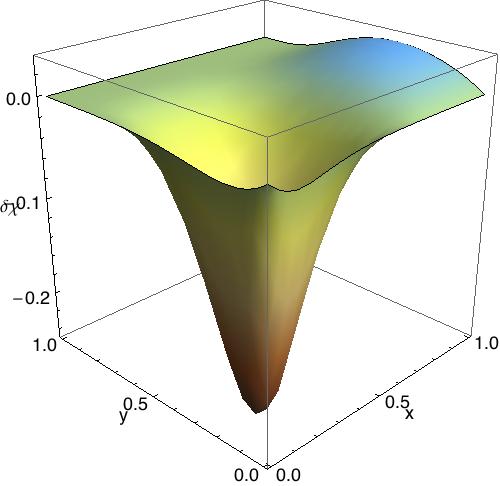}
\label{subfig:chil2} } \subfigure[]{
\includegraphics[scale=0.30]{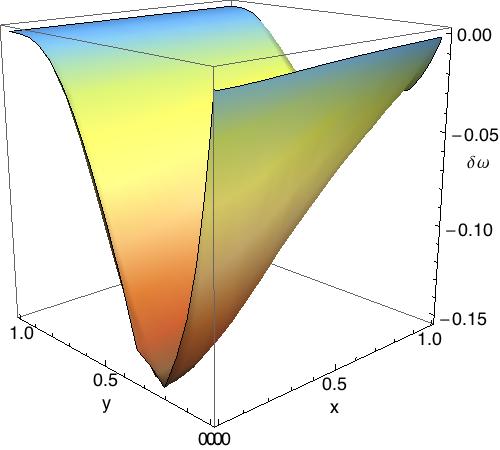}
\label{subfig:omegal1} } \hspace{0.5cm} \subfigure[]{
\includegraphics[scale=0.30]{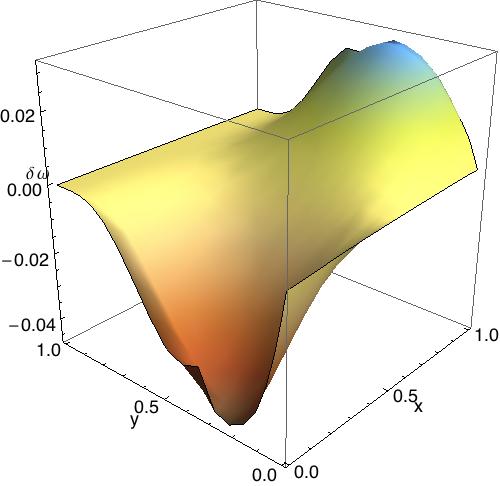}
\label{subfig:omegal2} }
\end{center}
\caption{Functions $\delta\chi(x,y)$ and $\delta\omega(x,y)$  for
the $\ell=1$ (Figs. \ref{subfig:chil1} and \ref{subfig:omegal1}) and
$\ell=2$ (Figs.  \ref{subfig:chil2} and \ref{subfig:omegal2}) modes
respectively.} \label{fig:perturbations2}
\end{figure}

For  $a/r_m>(a/r_m)_\textrm{mem}$, i.e., in the ultraspinning
regime, there  appears an infinite sequence of new negative modes.
The values of $a/r_m$ at which the $\ell=2,3,4$ branches intersect
$k_c=0$ are displayed in Table \ref{Table:critRot}.  These modes do
not admit a thermodynamic interpretation and therefore they should
correspond to new perturbative black holes with deformed horizons
(see points $A,B,C$ in Fig. \ref{fig:phases}). In particular, the
$\ell=2$ mode should signal the onset of the ultraspinning
instability conjectured in \cite{Emparan:2003sy}. For $d=7$ the
actual metric perturbations
${\delta\mu_0,\,\delta\mu_1,\,\delta\chi,\,\delta\Phi}$ for the
$\ell=1,2$  modes are displayed in Figs.
\ref{fig:perturbations1}-\ref{fig:perturbations2}. Figures
\ref{subfig:mu0l1}, \ref{subfig:mu0l2}, \ref{subfig:mu1l1} and
\ref{subfig:mu1l2} show that for each of the $\ell=1,2$ modes, the
number of zeros that $\delta\mu_0(x,y)$ and $\delta\mu_1(x,y)$ have
on the horizon $(y=0)$ coincides with the integer $\ell$. We have
checked that the same is also true for the higher harmonics with
$\ell>2$.

\begin{figure}[t!]
\begin{center}
\includegraphics[scale=0.6]{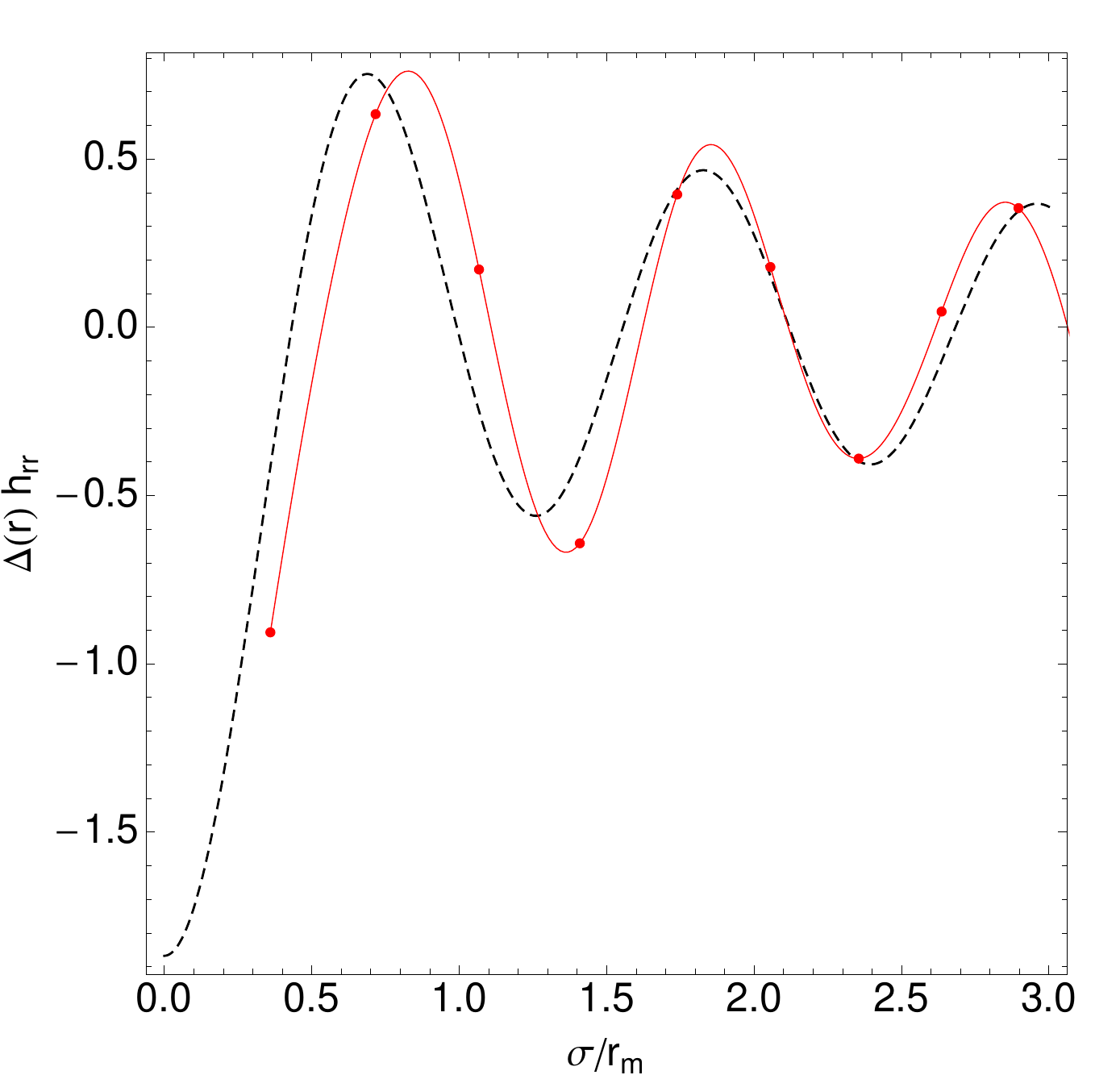}
\end{center}
\caption{Comparison between our $\Delta(r)\,h_{rr}$ for $a/r_m=4.38$ and $k_c^2r_m^2=0.114$ in $d=7$ (red dots) and the fitting function $\alpha\,J_0(\kappa\,\sigma)$ (dashed line). The red line corresponds the interpolating polynomial to our data and should only serve to guide the eye. To obtain the plots, the minimum value of $x$ was taken to be $x_\textrm{min}=0.69804$, since the comparison should be relevant near the rotation axis at $x=1$.}
\label{fig:bessel}
\end{figure}

According to the discussion in subsection
\ref{subsec:ultramotivations}, an important prediction  of
\cite{Emparan:2003sy} is that, in the $a\to \infty$ limit
\textit{and} in a region close to the axis $\theta\approx 0$, the
threshold axisymmetric modes at the horizon should be
well-approximated by a Bessel function $J_0(\kappa\,\sigma)$. In
order to check if, for large $a$ and close to the horizon and the
axis, our perturbations reduce to a Bessel function (at least
qualitatively), we have fitted our numerical results in $d=7$ for
$\Delta(r)\,h_{rr}$ with $a/r_m=4.38$ and
$k_c^2r_m^2=0.114$,\footnote{For these results $y=0.0034$.} with a
Bessel function:
\begin{equation}
\alpha\,J_0(\kappa\,\sigma)\,,\quad \textrm{with}\,\quad
\alpha=-1.868\,,\;\;\kappa=5.564\,,
\end{equation}
and $\sigma=a\,\sin\theta$. The results are depicted in
Fig.~\ref{fig:bessel} and they show that the agreement is  quite
remarkable. We should emphasise though that the argument of
\cite{Emparan:2003sy} only applies in the strict limit mentioned
above. For our data $a/r_m$ is relatively large (compared to the
onset of the ultraspinning instability) but nevertheless finite, and
therefore only a qualitative agreement with the prediction of
\cite{Emparan:2003sy} should be expected. This is what
Fig.~\ref{fig:bessel} shows.

\begin{figure}[t]
\centering \subfigure[]{
\includegraphics[width =4.8 cm]{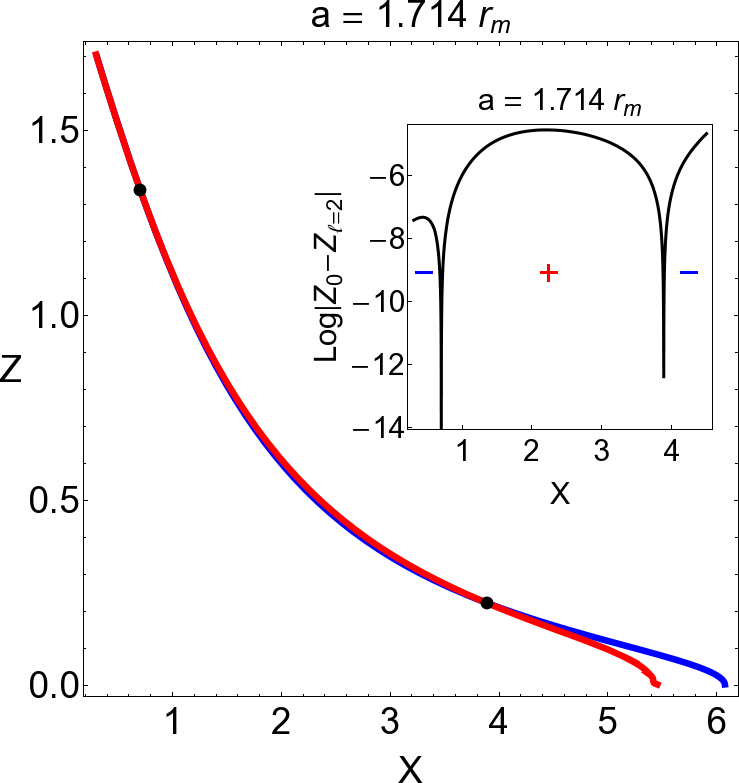}
\label{fig:Esubfig1} } \hspace{0.2cm} \subfigure[]{
\includegraphics[width = 4.8 cm]{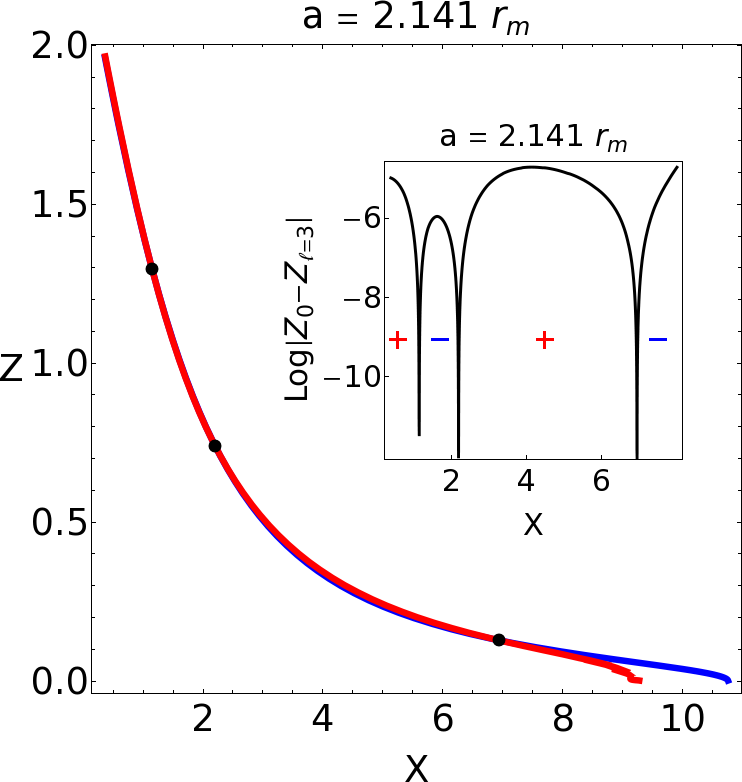}
\label{fig:Esubfig2} } \hspace{0.2cm} \subfigure[]{
\includegraphics[width = 4.8 cm]{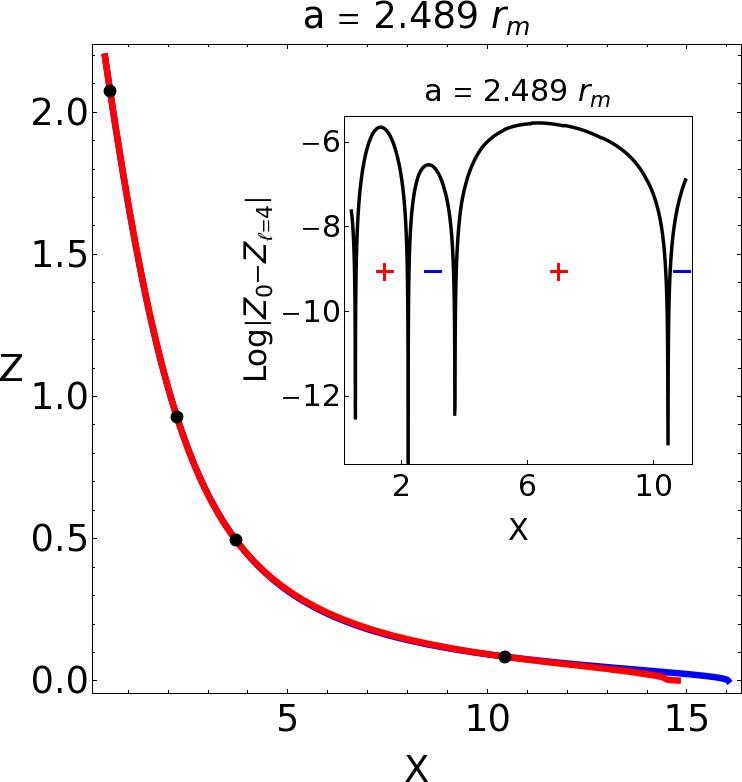}
\label{fig:Esubfig3} }
\\
\vspace{0.2cm} \subfigure[]{
\includegraphics[width = 4.8 cm]{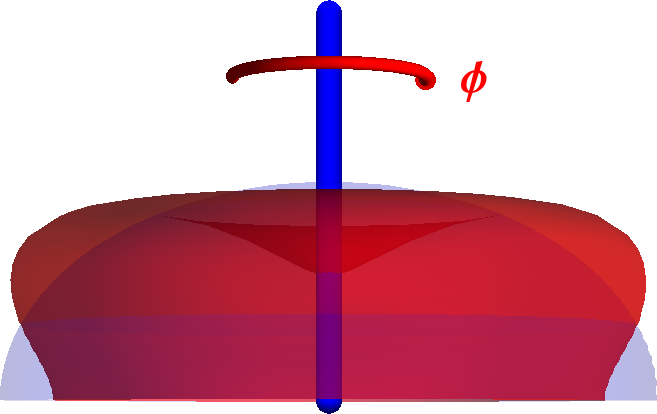}
\label{fig:Esubfig4} } \hspace{0.2cm} \subfigure[]{
\includegraphics[width = 4.8 cm]{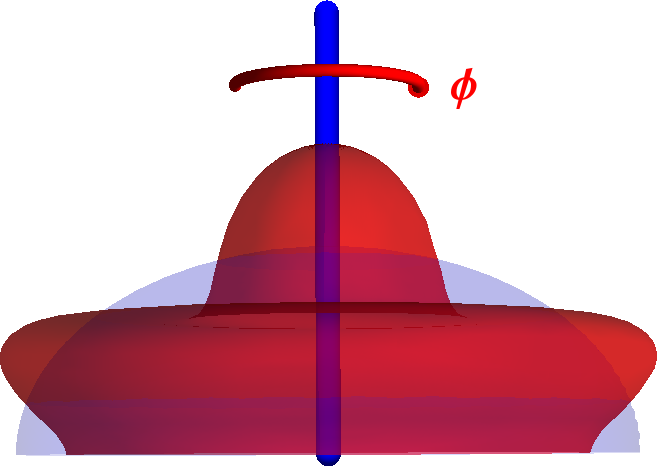}
\label{fig:Esubfig5} } \hspace{0.2cm} \subfigure[]{
\includegraphics[width = 4.8 cm]{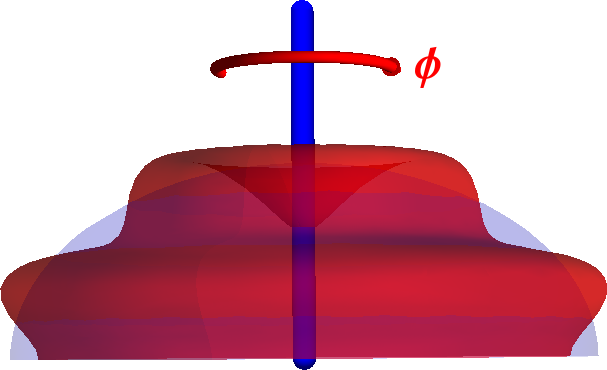}
\label{fig:Esubfig6} } \caption{ Embedding diagrams (Figs.\ref{fig:Esubfig1},\ref{fig:Esubfig2}, \ref{fig:Esubfig3}) at $(a/r_m)_\mathrm{crit}$ of the $d=7$ black hole horizon, unperturbed (blue), and perturbed (red) with the $\ell=2,3,4$ harmonics (in this order). The embedding Cartesian coordinates $Z$ and $X$ lie along the rotation axis $\theta=0$ and the rotation plane $\theta=\pi/2$ respectively. We also show the logarithmic difference between the embeddings of the perturbed ($Z_{\ell=2,3,4}$) and unperturbed ($Z_0$) horizons. The number of spikes corresponds to the number of crossings between the two embeddings. Each red $+$ sign indicates that the perturbation bulges out relatively to the background, and the blue $-$ signs indicates the opposite situation. Figures \ref{fig:Esubfig4}, \ref{fig:Esubfig5} and \ref{fig:Esubfig6} provide a three-dimensional illustration of the spatial sections of the horizon, with the $(d-4)$-sphere suppressed at every point, with the same color code.}
\label{fig:embeddings}
\end{figure}

To confirm our interpretation and visualise the effect of these
perturbations on the horizon of the background MP black hole, in
Fig.~\ref{fig:embeddings} we compare the embedding diagrams of the
unperturbed and the perturbed horizons in $d=7$ (see Appendix
\ref{app:embeddings} for the technical details of the embedding
diagram construction). For other values of $d$ the picture is
qualitatively similar. From the embeddings we conclude that the
effect of the $\ell=2$ modes is to create a pinch centered on the
axis of rotation;  the $\ell=3$ modes create a pinch at a finite
latitude and  the $\ell=4$ modes create two pinches, one centered on
the rotation axis and the other at a finite latitude. These are
precisely the kind of deformations depicted in
Fig.~\ref{fig:phases}.

Finally we notice that the value of $(a/r_m)_\textrm{crit}$  for every $\ell>1$ increases with the number of spacetime dimensions $d$. It would be interesting to explore if the perturbative approach of \cite{Emparan:2009at} can capture the dynamics of these instabilities in the  $a\to \infty$ limit.

%%%%%%%%%%%%%%%%%%%%%%%%%%%%%%%%%%%%%%%%%%%%%%%%%%%%%%%%%%%%%%%%%%%%%%%%%%%%
\section{Discussion}
\label{sec:discussion}
%%%%%%%%%%%%%%%%%%%%%%%%%%%%%%%%%%%%%%%%%%%%%%%%%%%%%%%%%%%%%%%%%%%%%%%%%%%%

In this paper we have studied in detail the properties of the onset of the ultraspinning instability in singly-spinning MP black holes in $d\geq 6$. This instability is captured by a class of perturbations that preserve the $\mathbb R \times U(1)\times SO(d-3)$ symmetries of the background, as well as the angular velocity and temperature of the original MP black hole. These perturbations (in the TT gauge) must satisfy the Lichnerowicz system of equations \eqref{Lichnerowicz}. Our strategy however was to solve the more general {\it eingenvalue}  problem \eqref{eigenh} because: i) there are powerful numerical routines to solve generalized eigenvalue problems of this kind; and ii) this also allows us to study Gregory-Laflamme$-$like instabilities of the associated black string. In short, varying the rotation parameter $a$ of the background black hole \eqref{mpbh} we searched for the negative modes of the problem \eqref{eigenh}, \ie for solutions that have generically $k_c\neq 0$. We found several families of negative modes that exhibit an underlying harmonic structure, although the equations that we solve do not seem to separate. We used this to suggestively label the several branches by an integer $\ell$. This integer coincides with the number of zeros of the metric perturbations on the horizon. The family $\ell=0$ always has $k_c\neq 0$, and coincides with the well-known negative mode of the Schwarzchild-Tangherlini solution when the rotation parameter $a$ vanishes \cite{Gross:1982cv,Kudoh:2006bp}. This family therefore describes the critical mode $|k|=k_c$ of the original Gregory-Laflamme instability of the black string \cite{Gregory:1993vy} when $a=0$, and our results show how it evolves as the rotation increases: the value of the threshold wavenumber $k_c$ increases with the rotation.

The branches with $\ell\geq1$ are more interesting since they intersect the $k_c=0$ axis at a critical $a>0$ (see Table \ref{Table:critRot}). Thus they describe not only {\it new} types of Gregory-Laflamme instabilities of rotating black strings, but they also represent, for $\ell\geq2$, true instabilities of the MP {\it black hole}. For black strings,  the onset of these instabilities is conjectured to signal a bifurcation to new branches of non-uniform black strings in which the horizon is deformed along both the direction of the string \textit{and} the polar direction of the transverse sphere. At a given rotation, the $\ell=1$ mode has the shortest wavelength $k_c^{-1}$ and hence it should dominate the instability of the black string.

From now on, we focus our attention only on the consequences of our
findings for the stability of the MP black holes.  The critical
rotation where the $\ell=1$ mode appears can be predicted by the
thermodynamic argument of \cite{Dias:2009iu,Dias:2010eu} (see
Section 2.2), which can be seen as a refinement of the Gubser-Mitra
conjecture. Accordingly, this mode  does {\it not} correspond to a
true instability of the MP black hole. Instead, we have interpreted
it as a thermodynamic mode that corresponds to a variation of the
parameters within the MP family of solutions such that it preserves
the temperature and angular velocity of the background (Section
\ref{sec:results}, and point $0$ in Fig. \ref{fig:phases}).

The modes with $\ell\geq 2$ describe the onset of true ultraspinning
instabilities of the MP black hole.  In section \ref{sec:results},
we have studied the deformations that these modes produce on the
horizon of the black hole for $d=7$ (just for concreteness) and
found that they give rise to the kind of deformations predicted in
\cite{Emparan:2007wm}. In addition, we have shown that, for large
$a$ and near $\theta=0$, our numerical results are well-approximated
by a Bessel function, which is in agreement with the heuristic
arguments of \cite{Emparan:2003sy}. Altogether, these results
provide solid evidence in favor of the ultraspinning instability
conjectured in \cite{Emparan:2003sy,Emparan:2007wm}. The thresholds
of the ultraspinning instabilities are expected to signal
bifurcations to new branches of axisymmetric solutions with pinched
horizons in a phase diagram of stationary solutions (see points
$A,B,C,...$ in Fig.~\ref{fig:phases}). Although these pinched black
holes have the same isometries as the original MP black hole, their
spherical-topology horizons are distorted by ripples along the polar
direction. When continued in the full non-linear regime, these new
branches of solutions  are conjectured to connect to the black ring
and black Saturn families \cite{Emparan:2007wm}. Although we have
identified the critical values of the rotation where these new
branches of pinched black holes appear, unfortunately with our
methods we cannot determine if they will have larger or lower
entropy in the phase diagram $S$ vs. $J$ at fixed total mass. This
would require going beyond linear order in perturbation theory.

We have limited our study to \textit{stationary} perturbations and
therefore we  cannot claim to have found an unstable mode, i.e., a
linear perturbation (satisfying our boundary conditions) that grows
exponentially with time; we have only found the stationary
zero-modes that  signal the onset of the instability. Including
time-dependence is not conceptually difficult but it is only
technically harder. As pointed out in footnote \ref{foot:timedep},
consistency of \eqref{eigenh} requires turning on extra components
of the perturbed metric.  This problem is however of fundamental
interest since its solution would provide a definite proof of the
ultraspinning instability together with information on its
timescale. The analogous stability problem, including
time-dependence, was studied in \cite{Dias:2010eu} in the context of
odd-dimensional MP black holes with equal angular momenta in all
rotation planes. There, the analogue of the $\ell=1$ zero-mode is
also present, and the time-dependent analysis confirmed the absence
of a black hole instability in this sector. We take this result as
good evidence in support of a similar interpretation in the
singly-spinning case.

The $d=5$ case is special. As \cite{Emparan:2003sy} already pointed
out, the $d=5$ singly-spinning MP black hole  does not have an
ultraspinning regime (in the sense of subsection
\ref{subsec:ultrathermo}) and therefore a priori there is no
argument that suggests the existence of an instability within the
class of perturbations that preserve the isometries of the
background. In this paper we have confirmed this picture since in
$d=5$ we only found one negative mode (the $\ell=0$ branch) which
has $k_c\neq 0$ for all values of $a$ such that $a<r_m$. The $d=4$
case has already been discussed in \cite{Monteiro:2009ke}.

An interesting open question to be addressed is whether the ultraspinning instability is also present in MP black holes in AdS. This background might introduce new features and physics that deserve a study. In particular, AdS might exhibit new black hole phases with an interesting interpretation in the holographic dual field theory. In fact, pinched plasma balls \cite{Lahiri:2007ae}, new kinds of deformed plasma tubes \cite{Cardoso:2006ks} and rotating plasma ball instabilities \cite{Cardoso:2009bv} have been found in the context of Scherk-Schwarz-AdS. The spectral methods that we have employed can also be applied in AdS spacetimes (see \cite{Monteiro:2009ke} for an application in $d=4$).

The existence of an ultraspinning regime is not unique to MP black holes. For instance, the five-dimensional black rings also admit a certain ultraspinning regime.  Namely, in the limit of arbitrarily large angular momentum the black ring resembles a boosted black string \cite{Emparan:2004wy}, which is known to suffer from GL instabilities \cite{Hovdebo:2006jy}. Moreover, the arguments in \cite{Emparan:2001wn,Hovdebo:2006jy,Elvang:2006dd} suggest that black rings should suffer from this GL instability for relatively low values of the angular momentum. Although these GL modes would break the symmetries of the background, it is interesting to ask whether there are zero-modes which can be captured by our methods. In fact, black rings are expected to suffer from other types of instabilities. For instance, fat rings are expected to be unstable under changes of their radius \cite{Elvang:2006dd}, and doubly-spinning black rings are conjectured to suffer from similar instabilities as well as from superradiant instabilities associated to the rotation along $S^2$ \cite{Dias:2006zv}.

Rotating asymptotically flat black holes can also suffer from
instabilities which break the axisymmetry.  Recently,
\cite{Shibata:2009ad,Shibata:2010wz} performed a remarkable fully
non-linear numerical evolution of a perturbed singly-spinning MP
solution in $d\geq 5$ and found that these black holes are
dynamically unstable against {\it non}-axisymmetric bar-mode
perturbations. The endpoint of this instability (at least in the
regime of rotations explored in \cite{Shibata:2010wz}) seems to be
another MP black hole with a smaller angular momentum. Notice
however that, because the axisymetry is broken, the threshold of the
instability is not expected to be stationary, and thus there will be
no bifurcation to a new stationary family of black holes. The
existence of such an instability had been conjectured also by
Emparan and Myers in \cite{Emparan:2003sy} using a thermodynamic
argument of a different type to the one explored in the current
manuscript.\footnote{The original argument of \cite{Emparan:2003sy}
for the bar-mode instability is the following. The entropy of a
highly rotating black hole can be smaller than that of two boosted
Schwarzchild-Tangherlini black holes in orbital motion. Therefore it
should be entropically favored to the system to develop a
non-axisymmetric configuration.} This bar-mode instability is
certainly different in nature from the axisymmetric instability,
which is not present in the $d=5$ case. Moreover, the bar-mode
instability becomes active at slightly lower rotations: compare the
critical values of $a/r_m$ for the onset of the bar instability in
Table 1 of \cite{Shibata:2010wz} with the critical values of $a/r_m$
for the onset of the ultraspinning instability in our Table
\ref{Table:critRot}. It would be certainly very interesting to study
also the non-linear time evolution of the axisymmetric ultraspinning
instability.

In addition, in the presence of certain matter or a cosmological constant, rotating or charged black holes (including the Kerr solution) can be unstable due to the superradiant phenomena \cite{Kunduri:2006qa,BHbomb,Cardoso:2006wa,superradiance}. Consider a black hole with angular velocity $\Omega_H$ (or with chemical potential $\mu_H$). If the frequency $\omega$ and the angular momentum $m$ (or charge $e$) of a scattering wave satisfy the relation $\omega<m\Omega_H$ (or $\omega<e\mu_H$), the scattering is superradiant: the wave extracts rotational energy (or electromagnetic energy) from the background and gets amplified. The presence of an effective reflective potential barrier, which may be due to the mass in a massive scalar field case or due to the AdS gravitational potential, combined with superradiance can then lead to multiple wave amplification and reflection that renders the system unstable. This can occur for scalar, electromagnetic or gravitational (but not fermionic) waves. Such an unstable system is often coined as a ``black hole bomb" \cite{BHbomb}. It would be very interesting to determine  the endpoint of this superradiant instability \cite{Kunduri:2006qa,Cardoso:2006wa} and there is active ongoing research in this direction \cite{Basu2010}. The spectral methods might also be useful to tackle this problem.

Returning back to Smarr's expectation of an instability in rotating black holes based on an analogy between rotating black holes and fluid droplets \cite{Smarr:1973zz}, it is interesting that it is realised in higher dimensions with the ultraspinning instability. That this happens is no longer seen as a mere analogy but a consequence of the gauge/gravity correspondence.  Black holes are thermodynamic objects and as such one should expect that, in a long wavelength regime (compared to the energy scale set by the temperature), they should have an effective hydrodynamic description. The first attempt to materialize this idea was the membrane paradigm \cite{membrane}.  More recently, the AdS/CFT duality motivated further research in this direction which culminated with the formulation of  the fluid/gravity correspondence \cite{Bhattacharyya:2008jc}, which provides a precise hydrodynamic description of asymptotically large AdS black holes. In addition, we should mention the blackfold approach of \cite{Emparan:2009at,Emparan:2009vd} which also provides a hydrodynamic description of \textit{asymptotically flat} black holes in the regime where this approximation applies. In both latter cases, the formal connection between gravity and fluid dynamics is established through a derivative expansion of the Einstein equations.

In this context, it is therefore with no surprise that many of the aforementioned black hole instabilities have an hydrodynamic description. The Gregory-Laflamme instability \cite{Gregory:1993vy} of black strings is in correspondence with the Rayleigh-Plateau instability  in fluid tubes \cite{Cardoso:2006ks} and with damped unstable sound wave oscillations \cite{Camps:2010br}. The subject of our study, the ultraspinning instability and the associated new phases of pinched stationary black holes, also possesses a fluid description as ultraspinning pinched plasma balls \cite{Lahiri:2007ae}. Finally, the bar mode non-axisymmetric instability \cite{Shibata:2009ad} was also conjectured to exist in black holes due to its presence in rotating fluids \cite{Cardoso:2009bv}.

%%%%%%%%%%%%%%%%%%%%%%%%%%%%%%%%%%%%%%%%%%%%%%%%%%%%%%%%%%%%%%%%%%%%%%%%%%
\section*{Acknowledgements}
%%%%%%%%%%%%%%%%%%%%%%%%%%%%%%%%%%%%%%%%%%%%%%%%%%%%%%%%%%%%%%%%%%%%%%%%%%%%

It is a pleasure to acknowledge the stimulating discussions with our collaborators Roberto Emparan and Harvey Reall in the ultraspinning project. OJCD acknowledges financial support provided by the European Community through the Intra-European Marie Curie contract PIEF-GA-2008-220197. PF is supported by an STFC rolling grant. RM and JES acknowledge support from the Funda\c c\~ao para a Ci\^encia e Tecnologia (FCT-Portugal) through the grants SFRH/BD/22211/2005 (RM) and SFRH/BD/22058/2005 (JES). This work was partially funded by FCT-Portugal through projects PTDC/FIS/099293/2008, CERN/FP/83508/2008 and CERN/FP/109306/2009.

%%%%%%%%%%%%%%%%%%%%%%%%%%%%%%%%%%%%%%%%%%%%%%%%%%%%%%%%%%%%%%%%%%%%%%%%%%
\appendix

%%%%%%%%%%%%%%%%%%%%%%%%%%%%%%%%%%%%%%%%%%%%%%%%%%%%%%%%%%%%%%%%%%%%%%%%%%%%
\setcounter{equation}{0}
\section{Embedding diagrams}
\label{app:embeddings}
%%%%%%%%%%%%%%%%%%%%%%%%%%%%%%%%%%%%%%%%%%%%%%%%%%%%%%%%%%%%%%%%%%%%%%%%%%%%

In this appendix, we summarise the main ingredients of the embeddings presented in Fig. \ref{fig:embeddings}. Since the kind of perturbations we are considering preserve the transverse $(d-4)$-sphere, we will henceforth suppress it and concentrate on the non-trivial four-dimensional part of the metric.  In order to be able to visualise the geometry of the horizon for \textit{all} values of the rotation parameter we will adopt the embedding proposed in \cite{Frolov:2006yb}  in the context of the Kerr-Newman black hole.

Recall that for both the perturbed and the unperturbed geometries, the spatial cross-sections of the future event horizon $\mathcal H^+$ are hypersurfaces of constant $t$ and $r=r_+$.  Here $r_+$ denotes the location of the event horizon of the background MP black hole. The induced metric on this hypersurface (suppressing the transverse $(d-4)$-sphere) can be written as
 \beq\label{emb:2d}
ds^2_{2d}= V(\sigma) d\sigma^2 +\sigma^2 d\phi^2\,, \qquad
\hbox{with} \quad \sigma=\sqrt{g_{\phi\phi}}{\bigr |}_{r_+}\,, \quad
\hbox{and} \quad V(\sigma)=g_{xx}{\bigr |}_{r_+}
\lp\frac{dx}{d\sigma}\rp^2\,,
 \eeq
where $x=\cos\theta$. This two-dimensional surface of revolution can be embedded into Euclidean four-dimensional space $\mathbb{E}^4$ via the map $(\sigma,\phi)\mapsto \lp X,Y,Z,R \rp$ defined by
 \beq
 \label{emb:map}
  \lp X,Y,Z\rp=\frac{\sigma}{\Phi_0}{\bigl(}
  F(\phi),G(\phi),H(\phi){\bigr)}\,,\quad R=R(\sigma)\,,\qquad
  \hbox{and}\quad F(\phi)^2+G(\phi)^2+H(\phi)^2=1\,,
 \eeq
and the induced metric for the associated 2-surface is
 \beq
 \label{emb:2dnew}
 ds^2_{\mathbb{E}^4}= dX^2+dY^2+dZ^2+dR^2
 = \left[ \Phi_0^{-2}+\lp\frac{dR}{d\sigma}\rp^2 \right] d\sigma^2 +\sigma^2
d\phi^2\,,
 \eeq
where we ask the reader to see \cite{Frolov:2006yb} for a detailed understanding of the second equality.

 Introducing the scale parameter $\eta$ and the distortion parameter
$\beta$,
 \beq \label{emb:parameters}
  \eta=\lp r_+^2+a^2 \rp^{1/2}\,, \quad \beta=a\lp r_+^2+a^2 \rp^{-1/2} \,,
 \eeq
it follows that
 \beq \label{emb:parameters2}
 g_{xx}{\bigr |}_{r_+}=\eta^2 f(x)\,, \quad g_{\phi\phi}{\bigr |}_{r_+}=\eta^2
 f(x)^{-1}\,, \qquad \hbox{with} \quad
 f(x)=\frac{1-x^2}{1-\beta^2(1-x^2)}\,.
 \eeq
Then, from $ds^2_{\mathbb{E}^4}=ds^2_{2d}$, we get
 \beq \label{emb:Rprime}
 R'(x)=\eta \,f(x)^{-1/2}\sqrt{1-\frac{f'(x)^{2}}{4\Phi_0}}\,.
 \eeq
Reality of this expression requires that $\Phi_0\geq |f'(x)|/2$ for
$0\leq x\leq 1$ which is satisfied if
  \beq \label{emb:Phi0}
 \Phi_0\geq \frac{1}{2}\, \max_{\,x\in [0,1]\,} \, |f'(x)| \,.
 \eeq

In the embedding diagrams shown in Fig. \ref{fig:embeddings}, the Cartesian coordinate $Z$ was chosen to be given by the $R(x)$ that solves \eqref{emb:Rprime}. Similarly, $X$ was chosen to be given by the function $f(x)^{1/2}$ defined in \eqref{emb:parameters2}.

%%%%%%%%%%%%%%%%%%%%%%%%%%%%%%%%%%%%%%%%%%%%%%%%%%%%%%%%%%%%%%%%%%%%%%%%%%

\end{document}